\documentclass[a4paper,oneside]{article}
\usepackage[margin=1in]{geometry}
\usepackage[numbers,sort&compress]{natbib}
\usepackage{amsfonts}
\usepackage{graphicx}
\usepackage{epstopdf}
\usepackage{algorithmic}
\ifpdf
  \DeclareGraphicsExtensions{.eps,.pdf,.png,.jpg}
\else
  \DeclareGraphicsExtensions{.eps}
\fi

\usepackage{mathrsfs}
\usepackage{amssymb}
\usepackage{amsmath}
\usepackage{mathtools}
\usepackage{subcaption}
\usepackage{xcolor,hyperref}
\usepackage{comment}
\usepackage[normalem]{ulem}
\usepackage{tikz}
\usetikzlibrary{calc,positioning,arrows.meta}
\usepackage{inputenc}
\usepackage{cleveref}
\usepackage{amsopn}

\usepackage{makecell}
\usepackage{tabularx}
\usepackage{booktabs}

\usepackage{enumitem}
\setlist[enumerate]{leftmargin=.5in}
\setlist[itemize]{leftmargin=.5in}

\newtheorem{remark}{Remark}

\crefname{hypothesis}{Hypothesis}{Hypotheses}

\def\d{{\partial}}
\def\D{{\mathcal{D}}}

\def\F{{\mathcal{F}}}
\def\g{{\mathbf{g}}}

\def\h{{\mathbf{h}}}

\def\lam{{\boldsymbol\lambda}}

\def\muu{{\boldsymbol\mu}}

\def\R{{\mathbb{R}}}

\def\u{{\mathbf{u}}}

\def\V{{\mathcal{V}}}

\def\x{{\mathbf{x}}}

\makeatletter
\newcommand*{\addFileDependency}[1]{%
  \typeout{(#1)}%
  \@addtofilelist{#1}%
  \IfFileExists{#1}{}{\typeout{No file #1.}}%
}
\makeatother

\newcommand{\revone}[1]{\textcolor{black}{#1}}
\newcommand{\revtwo}[1]{\textcolor{black}{#1}}
\newcommand{\revthree}[1]{\textcolor{black}{#1}}

\usepackage[mathlines]{lineno}
\newcommand*\patchAmsMathEnvironmentForLineno[1]{%
  \expandafter\let\csname old#1\expandafter\endcsname\csname #1\endcsname
  \expandafter\let\csname oldend#1\expandafter\endcsname\csname end#1\endcsname
  \renewenvironment{#1}%
     {\linenomath\csname old#1\endcsname}%
     {\csname oldend#1\endcsname\endlinenomath}}%
\newcommand*\patchBothAmsMathEnvironmentsForLineno[1]{%
  \patchAmsMathEnvironmentForLineno{#1}%
  \patchAmsMathEnvironmentForLineno{#1*}}%
\AtBeginDocument{%
\patchBothAmsMathEnvironmentsForLineno{equation}%
\patchBothAmsMathEnvironmentsForLineno{align}%
\patchBothAmsMathEnvironmentsForLineno{flalign}%
\patchBothAmsMathEnvironmentsForLineno{alignat}%
\patchBothAmsMathEnvironmentsForLineno{gather}%
\patchBothAmsMathEnvironmentsForLineno{multline}%
}
\title{Causal Multi-fidelity Surrogate Forward and Inverse Models for ICF Implosions}
\author{
Tyler E. Maltba \thanks{Theoretical Division, Los Alamos National Laboratory, Los Alamos, NM 87545, USA 
(\url{tyler.maltba@lancium.com}, \url{southworth@lanl.gov}, \url{mklasky@lanl.gov})}
\and Ben S. Southworth \footnotemark[2]
\and Jeffrey R. Haack \thanks{Computer, Computational, and Statistical Sciences Division, Los Alamos National Laboratory, Los Alamos, NM 87545, USA (\url{haack@lanl.gov})}
\and Marc L. Klasky \footnotemark[2]
}

\begin{document}
\maketitle

\begin{abstract}
Continued progress in inertial confinement fusion (ICF) requires solving inverse problems relating experimental observations to simulation input parameters, followed by design optimization. However, such high-dimensional dynamic PDE-constrained optimization problems are extremely challenging or even intractable. It has been recently shown that inverse problems can be solved by only considering certain robust features. Here we consider the ICF capsule's deuterium-tritium (DT) interface, and construct a causal, dynamic, multifidelity reduced-order surrogate that maps from a time-dependent radiation temperature drive to the interface's radius and velocity dynamics. The surrogate targets an ODE embedding of DT interface dynamics, and is constructed by learning a controller for a base analytical model using low- and high-fidelity simulation training data with respect to radiation energy group structure. After demonstrating excellent accuracy of the surrogate interface model, we use machine learning (ML) models with surrogate-generated data to solve inverse problems optimizing radiation temperature drive to reproduce observed interface dynamics. For sparse snapshots in time, the ML model further characterizes the most informative times at which to sample dynamics. Altogether we demonstrate how operator learning, causal architectures, and physical inductive bias can be integrated to accelerate discovery, design, and diagnostics in high-energy-density systems.
\end{abstract}

\section{Introduction}

Notwithstanding the 2022 breakthrough demonstration of ignition at the National Ignition Facility
(NIF), the more recent experiments at NIF reiterate (yet again!) the central role better modeling, diagnostics, and accompanying methods for design optimization and parameter inference from experimental observations will continue to play to ensure continued progress in Inertial
Confinement Fusion (ICF). Ensuring continued progress demands the ability to generate high-fidelity simulation data to address high-dimensional design optimization and parameter estimation problems necessary to achieve shot-to-shot reproducibility within a small range of uncertainty (5-10\% range) for an implosion where a significant fraction of the energy release is from alpha particle heating. Furthermore, continued progress demands better understanding of the interactions between the numerous physical models that govern the radiation-hydrodynamics (rad-hydro) and burn-physics of ICF to enable improvements in  predictive simulation capability.  Inevitably, gaining this understanding requires solving inverse problems in which experimental observations are related to simulation input parameters governing system evolution, followed by design optimization.

To address both design optimization and parameter estimation, the ICF community has largely utilized Bayesian optimization (BO), e.g., \cite{nora2017ensemble,humbird2018deep,peterson2017zonal,humbird2019transfer,hatfield2019blind}. This optimization procedure is constructed to enable a map of input parameters characterizing the design of an ICF capsule and laser settings to scalar output parameters such as ICF implosion neutron yield and X-ray diagnostics. To perform this optimization, surrogate models using either Gaussian processes (GPs) or deep neural networks (NNs) have been utilized [ibid]. However, the surrogate models that have been constructed are generated in a manner in which the rich spatio-temporal inputs (e.g., 2D and 3D topologies of the ICF capsule, laser drive history, etc.) as well as spatio-temporal outputs are reduced to simple summary indicators and/or hand-engineered features such as the integral of an image, the peak of a time history, or the width of a spectral line. Such an approach limits the effectiveness of the entire analysis chain as most information from both experiments and simulations is either highly compressed or entirely ignored \cite{anirudh2020improved}. Unsurprisingly, surrogate models designed to predict these features are often under-constrained, ill-conditioned, not very informative, and overall insufficient to elucidate complex ICF physics [ibid]. \revtwo{Another troubling aspect of these procedures, particularly with respect to surrogate forward models regarding both globally aggregated parameters (e.g., taking averages of time histories) and network architectures (e.g., vanilla/traditional feedforward and convolutional NNs), is that they are not \revthree{inherently} causal, which further restricts their utility in elucidating the underlying physics of time-ordered, dynamic ICF systems. Causality in this context refers to the principle that current and past states of a dynamic system determine its future states, ensuring a clear temporal precedence where an effect cannot occur before its cause. Additional key aspects of causality include the existence of a functional/mechanistic relationship between inputs and outputs, bounded propagation of causal efficiency (e.g., finite mass-energy transfer velocity), predictability, and intervention, i.e., changing a cause/input should predictably change the effect/output. Hence, a time-ordered output prediction at a fixed time (e.g., implosion dynamics at $t=s$) should depend only on its time-dependent input(s)/source(s) up to that time (e.g., laser drive at $t\le s$), i.e., there is no looking forward in time.} Without causality, forward surrogates are entirely non-interpretable and non-physical. Finally, surrogate forward models are often inconsistent with the inverse, leading to an implausible overall system in which the intuitive cycle of mapping inputs to outputs and back to inputs can produce wildly varying results. Not only can an inverse prediction from the surrogate output be far away from the initial input, but even univariate sensitivities, i.e., inferred changes in predictions with respect to a single scalar input parameter, are often unintuitive [ibid].

A few attempts to overcome non-causal machine learning (ML) models as well as the highly limited dimensionality of input parameters have been made. However, in these attempts, the data utilized was not from simulation codes but rather from semi-analytical models to provide vast amounts of data for training the proposed causal NNs \cite{anirudh2020improved,humbird2018predicting}. This may be compared to the largest simulation data set utilized to train non-causal ML models, consisting of $6\times10^4$ samples, which consumed 39 million CPU hours of simulation time and only explored 9 input parameters. This underscores the need to speed up simulations to enable higher dimensional problems to be examined \cite{gaffney2014thermodynamic}.

To reduce the computational demands in developing training data for GP-based BO, multi-fidelity (MF) learning has been explored. Traditionally, ICF design has relied on low-fidelity (LF) modeling to initially identify potentially interesting design regions, which are then subsequently explored via selected high-fidelity (HF) modeling \cite{vazirani2021coupling}. However, it has recently been observed that this two-step approach can be insufficient: even for simple design problems, a two-step optimization strategy can lead HF searching towards incorrect regions and consequently waste computational resources on parameter regimes far away from the true optimal solution due to the presence of LF optima in distinct regions of the parameter space far from HF optima. To address this issue, an iterative MF BO method based on GP Regression that leverages both low- and high-fidelity modeling was proposed \cite{wang2024multifidelity}. However, this method utilizes a pre-trained non-causal surrogate forward model for data generation and makes assumptions regarding the ability to combine low- and high-fidelity simulations to navigate to an optimal design. \revthree{Building on this direction, recent MF BO work has made progress on ICF design optimization. Vazirani \textit{et al.}~\cite{vazirani2023coupling} coupled LF (1D xRAGE) and HF (2D xRAGE) simulations in a GP-based MF BO framework to optimize a four-parameter graded-inner-shell design, identifying double-shell capsule configurations with approximately 20\% higher no-burn DT yield than a baseline bilayer using $\sim$35 HF simulations. More recently, Peterson \textit{et al.}~\cite{peterson2024icecap} (project ICECap) developed an integrated MF BO workflow that scales to higher-dimensional ICF design: a 17-parameter 1D hohlraum study converged to a high-yield design within a few hundred equivalent LF simulations, and the resulting parameter adjustments (longer second shock, thicker ablator, smaller hohlraum, thinner DT ice) align with the manually-engineered design modifications that produced ignition and target gain at the NIF. These works collectively demonstrate that MF BO can reduce HF simulation cost and scale to ICF design spaces of moderate dimensionality.}

\revone{Outside of ICF, recent operator-learning surrogates—including neural operators and their transformer variants—have pushed accuracy and generalization on PDE benchmarks, but typically learn non-causal input-to-field maps without an explicit embedded low-dimensional dynamics model (e.g., see Neural Operator (NO)/JMLR survey \cite{kovachki2023neural}, convolutional NOs \cite{raonic2023convolutional}, and transformer-based NOs \cite{shih2025transformers}).} \revone{Differentiable-physics approaches (e.g., physics-informed NO \cite{li2024physics}) inject PDE structure via soft constraints/residual penalties yet still optimize direct field predictors rather than a controller for a reduced ODE, which is an embedded hard constraint. Projection/ROM methods such as Operator Inference (OPINF) provide principled reduced models and stability/closure tools, but are not usually framed as MF residual learning on a controller, nor do they demonstrate forward–inverse cycle consistency or learned sparse-time experimental design \cite{uy2023operator}. In ICF and high-energy-density physics (HEDP) specifically, recent works emphasize GP/BO surrogates and meta-BO for pulse/target optimization, or component surrogates (e.g., ML replacements for sub-models like NLTE \cite{kluth2020deep}), as well as data-driven predictors of direct-drive response---valuable but distinct from our causal ODE-embedded, LF $\to$ HF controller correction, and cycle-consistent forward–inverse formulation.}

\revone{Recent work shows that blending scarce HF data with abundant LF information can substantially improve neural operator training efficiency and accuracy across PDE surrogates. MF Deep Operator Network (DeepONet) variants demonstrate consistent gains in both purely data-driven and physics-informed settings by fusing LF and HF branches or learning HF--LF discrepancies \cite{Lu2022,Howard2023,BiFiDeepONetFluids2023}. In parallel, Fourier Neural Operator (FNO) approaches have been extended to MF settings, leveraging FNO’s resolution invariance to transfer across discretizations and reduce HF data requirements in large-scale scientific applications \cite{Tang2024MFFNO}. Wavelet Neural Operators have also been adapted to MF learning, where wavelet-domain coupling of fidelities yields efficient surrogates with quantified uncertainty and reliability guarantees \cite{Thakur2022MFWNO,Tripura2024MFWNOReliability}. To overcome limitations in both initial conditions and temporal causality, novel causal DeepONet architectures have been proposed \cite{nghiem2023causal,Liu2024CausalityDeepONet}. However, while these are promising directions for addressing causality in operator networks, their success on systems exhibiting complex dynamics such as those in ICF has not yet been demonstrated---a gap we seek to fill. Overall, these results indicate that MF operator networks provide a practical path to accurate surrogates when HF data are scarce, while remaining compatible with physics constraints and mesh-agnostic training.}

In summary, design optimization and parameter estimation for ICF are difficult due to the prohibitively expensive nature of the forward model. \revthree{While MF modeling in a BO setting has demonstrated meaningful computational savings and scaling to moderate dimensional ICF design optimization~\cite{vazirani2023coupling,peterson2024icecap}, these approaches---by construction---target scalar summary objectives (e.g., yield) as functions of static summary design inputs, using non-causal, non-dynamic surrogate or direct-simulation evaluations. As such, they are not structured to elucidate the time-ordered physical relationships---for example, between drive dynamics and implosion dynamics---that are needed to generalize outside
of a design baseline or to solve parameter-estimation inverse problems from time-resolved observations. The question of whether dynamic, causal, cycle-consistent MF surrogates can be constructed for ICF, and whether they can support such inverse problems, remains open.}

\revthree{These BO-based contributions are complementary to---but distinct from---our formulation in both goal and mechanism. For a BO design loop that terminates in an evaluation by the true simulator (e.g., xRAGE), surrogate cycle-consistency is not a prerequisite for the final design being physically consistent with the HF physics~\cite{peterson2024icecap}. Our setting differs: the forward surrogate is itself the artifact used downstream, both for inverse problems and for sparse-time experimental design, so empirical forward--inverse cycle consistency becomes a meaningful diagnostic of surrogate reliability.} \revtwo{In this context, cycle consistency involves solving for the ``optimized parameters" and then inserting them back into the hydro-dynamic solver or the surrogate model to demonstrate self-consistency.} \revthree{Relatedly, while BO surrogates can accurately map static inputs to scalar objectives, non-causality and the absence of an embedded dynamic model restrict their use in elucidating time-ordered physics---for example, relating time-dependent characteristics of the drive to specific dynamic observations of the implosion---which motivates our causal, ODE-embedded construction as a step toward physics-interpretable surrogates.}

\revone{Synthesizing these developments, three gaps persist—\emph{causality}, \emph{multi-resolution/multi-fidelity integration within the same physics model}, and \emph{interpretability}}. 

We address these \revone{``gaps"} with the following contributions:
\begin{itemize}
  \item \revone{\textbf{Causal forward surrogate.} A \emph{controller-in-the-loop ODE} surrogate that maps drive histories to interface dynamics while enforcing time causality.}
  \item 
  \revone
  {\textbf{Multi-resolution within one physics model.} A causal LF network predicts controller coefficients and a causal HF residual network corrects those coefficients on the \emph{controller} itself (LF $\to$ HF transfer), yielding HF-accurate trajectories with limited HF data and disjoint LF/HF data splits.}
  \item 
 \revone{\textbf{Forward--inverse self-consistency.} Cycle-consistency diagnostics (stability/cycle metrics) indicating cycle error is limited primarily by forward generalization.}
  \item 
  \revone{\textbf{Sparse-time inverse design.} Inverse models that \emph{learn} informative snapshot times and can optionally use velocity to improve diversity and accuracy.}

\end{itemize}
\revone{Table 1 summarizes the distinctions between our proposed ML framework and recent work on neural operators and traditional BO based optimization methods.}

\revone{To highlight our innovations, we examine a ICF inverse problem} in which the drive and equation of state (EoS) estimation, can be solved using a small number of robust features, e.g. \cite{serino2025physics,bell2025learning}. In our case, we consider the ICF capsule's deuterium-tritium (DT) interface as our feature of interest. To formulate  a causal, dynamic, MF reduced-order surrogate for the map $T_r\mapsto \x$, in \Cref{sec:surrogate}, where $T_r(t)$ denotes the time-dependent radiation temperature drive and $\x(t):= (R_i(t),V_i(t))$ the interface's radius and velocity dynamics, we first formulate an embedding of DT interface dynamics from full rad-hydro simulations (see \Cref{sec:data} for rad-hydro code and data generation) as a parameterized ordinary differential equation (ODE) over interface radius and velocity. Physical insight provides a base parameterized ODE, and a causal NN model is then trained on LF simulation data to learn a parameterization or ``controller'' of this ODE as a function of temperature drive, resulting in a surrogate LF forward model that maps temperature drive to a 2D ODE that can be rapidly integrated numerically to estimate DT interface dynamics. A second causal NN is then trained to perform a residual correction of LF network output to an ODE parameterization of HF DT interface data. The causal MF forward surrogate consists of the composition of these two networks, which is demonstrated to achieve high accuracy in reproducing DT interface dynamics with small amounts of HF training data in \Cref{sec:forward}.

Second, we use the MF surrogate to solve inverse problems related to estimating a target's external laser drive, in the form of a temperature source $T_r(t)$, from observed dynamics in \Cref{sec:param-nn}. In \Cref{subsec:dense-nn} we first consider estimating drives from entire interface trajectories using \revtwo{a sequence-to-static encoder NN, comprised of a Long Short-Term Memory (LSTM) network and attention pooling}, and demonstrate cycle consistency of the surrogate forward model from \Cref{sec:forward} and the inverse model. However, in experimental settings, only a small number of temporal snapshots of the capsule are available, and the specific times at which the snapshots are captured are design variables that are costly to tune. Thus, in \Cref{subsec:sparse-nn}, we develop a framework to simultaneously learn optimal discrete times at which to sample interface data, and then perform drive estimation from radius snapshots at those times. Although velocity is typically unavailable in experimental settings, to determine if this additional snapshot information improves drive estimation, we apply dynamic and global selection-based NNs using both radius and velocity snapshots. 

\newcommand{\Yes}{\checkmark}
\newcommand{\No}{\textemdash}
\newcommand{\tightcolspace}{\setlength{\tabcolsep}{4pt}}

\begin{table}[t]
\revone{
\centering
\tightcolspace
\begin{tabularx}{\linewidth}{l *{5}{c}}
\toprule
\makecell[l]{Approach\\(2023--2025)} & Causal & ODE-embed & MR/MF mech. & Fwd--Inv & Sparse-time \\
\midrule
Neural operators (DeepONet/NKN/Transformers) & \Yes & \No & \Yes & \No & \No \\
ICF ML/Optimization (GP/BO, data-driven)     & \No & \No & (co-kriging) & \No & \No \\
\textbf{This work (FMF + inverse)}            & \Yes & \Yes & \Yes & \Yes & \Yes \\
\bottomrule
\end{tabularx}
}
\caption{\revone{Contrasting representative recent approaches with this work along key dimensions for ICF forward and inverse modeling. ``Causal'' means dynamic outputs at time $s$ depend only on inputs at times $t\le s$. ``ODE-embed'' denotes a controller-in-the-loop surrogate rather than direct state prediction. ``MR/MF mech.'' indicates an explicit low $\to$ high mechanism operating on the controller. ``Fwd--Inv'' denotes empirical cycle-consistency diagnostics. ``Sparse-time'' denotes learned snapshot selection.}}
\label{tab:comparison}
\end{table}

\section{Data Generation}
\label{sec:data} 

Simulations were performed using Los Alamos National Laboratory’s
(LANL) xRAGE rad-hydro simulation code, which computes solutions in an Eulerian reference frame with adaptive mesh-refinement (AMR) \cite{Gittings_2008}. xRage is well benchmarked and widely used for the simulation of ICF and HEDP applications \cite{haines_layered_2017}. The hydrodynamics solver is a custom approximate Godunov-type solver for the Euler equations, similar to that of Harten-Lax-van Leer \cite{harten1983upstream}. The radiation transport equations are solved using a multi-frequency radiation diffusion approximation and a three-temperature (3T) plasma calculation \cite{winslow1995multifrequency}. Simulations employ LANL OPLIB opacity data, through the TOPS code, and LANL SESAME tabular EoS data \cite{colgan2016new,abdallah1985tops,lyon1978sesame}. Electron and ion thermal conductivities are based on the formulae of Lee and More with modifications \cite{lee1984electron,munro1994electron}.

\begin{figure}[h!]
    \centering

    \begin{subfigure}[t]{0.465\textwidth}
        \centering
        \includegraphics[width=\textwidth]{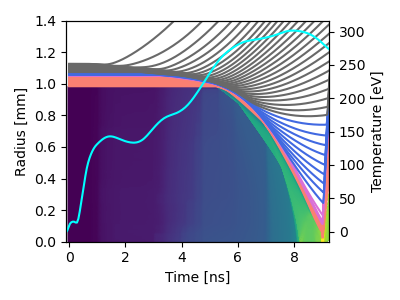}
        \caption{1D NIF shell configuration \revtwo{in $[mm]$} and simulated ICF implosion using 67 groups. 
        The shell is comprised of two outer layers of plastic (gray and blue) that are 
        ablated by the drive (cyan curve \revtwo{in $[eV]$}). These are followed by a very thin layer of 
        doped plastic (magenta), which serves to lower the adiabat of the DT ice (coral) 
        and aid in stable DT gas fill compression. The fill gas density evolution is 
        shown as a heat map.}
        \label{fig:implosion}
    \end{subfigure}
\quad
    \begin{subfigure}[t]{0.465\textwidth}
        \centering
        \includegraphics[width=\textwidth]{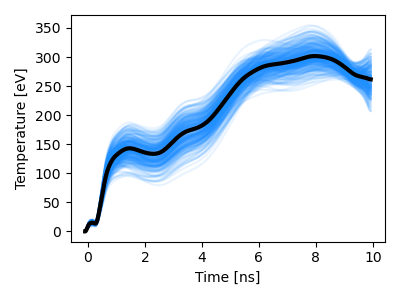}
        \caption{\revtwo{Visualization of the ensemble of drives in $[eV]$ used in this study. 
        The thicker, black line is the baseline drive that is modified to generate the ensemble.}}
        \label{fig:alldrives}
    \end{subfigure}
    \caption{(a) Implosion configuration and (b) ensemble of drives.}
\end{figure}

Our baseline simulation is inspired by NIF shot N221205.  This was a significant experiment conducted at NIF on December 5, 2022, which was groundbreaking because it achieved fusion ignition for the first time in a laboratory setting. We examine the impact of variations in the drive on the ``capsule-only'' implosions, as this aspect of the simulation is considered to be not only the most impactful on performance but also subject to the greatest uncertainty, leaving all other simulation parameters fixed \cite{vazirani2021coupling}. \Cref{fig:implosion} presents the baseline implosion dynamics for this investigation. The driving source imploding the capsule is modeled via a frequency-dependent source (FDS) in lieu of rigorous modeling of the hohlraum physics. This baseline FDS source was calculated using the HYDRA code and implemented as a  temperature flux $T_r(t)$ set on the boundary of the mesh \cite{shay2012implosion}.   Statistically independent simulations were run varying the drive and using both 3 \& 67 frequency groups.  We treat 3-group simulations as LF data and 67-group simulations as HF data in building our MF surrogate for dynamics of the DT interface, that is, the interface between the cryogenically frozen DT layer (i.e., DT ice) and DT gas fill. \revone{All simulations were performed using The Rocinante (“Roci”) cluster at Los Alamos National Laboratory. The system consists of 508 compute nodes, each with two Intel Xeon Platinum 8480 (Sapphire Rapids) CPUs for a total of 112 cores per node running at 2.0 GHz.  Simulations were performed using only 8 cores on a single node.  Average compute times were 15 minutes and 80 minutes for the three  and 67 group multi-group diffusion calculations.  Finally, computational savings of a factor of 25 was observed in 2D, emphasizing the heightened importance of multi-fidelity methods in  addressing higher-dimensional simulations.}

To generate new $T_r(t)$ drives, we first perform a spline fit of the nominal temperature drive using 35 knots. 
For each simulation, we generate a perturbation array for every other knot of the baseline spline fit by random numbers in the range $[-0.1,0.1]$, and add this to the original spline representation, then smooth the result by fitting to a spline at the original 35 knots. The associated FDSSs at each time are then adjusted to be consistent with the new radiation temperature. The result is a smooth drive, an example of which is provided in \Cref{fig:implosion} along with the resulting 1D ICF capsule implosion (via 67-group radiation diffusion). \revtwo{In \Cref{fig:alldrives}, we show an ensemble of drives generated by this process, with the thick line showing the `baseline' drive.}

\revtwo{The proposed surrogate is operator-valued and does not assume a fixed baseline, timing, or shock count. That is because our surrogate learns the operator $T_r(\cdot)\!\mapsto\!\x(\cdot)$ rather than a fixed design map, it is not restricted to a single baseline or a fixed number of shocks; generalization to different drive shapes and timings is governed by the diversity of the training distribution. In this study, the drive generator is restricted to $\pm10\%$ amplitude perturbations of a single baseline to isolate the MF (LF $\to$ HF) surrogate design and the forward–inverse pipeline. The same architecture and training procedure may be extended to broader drive families by (a) sampling multiple baselines with distinct foot/peak timings and numbers of shocks (via randomized knot sets and sub-pulse insertion/suppression in the spline generator), (b) applying amplitude scaling and time-dilation/warping augmentations, and (c) conditioning the controller and HF residual on a low-dimensional baseline descriptor (e.g., shock count or peg-point times). In all cases, the causal structure (outputs at time $s$ depend only on inputs with $t\!\le\! s$) and the controller-in-the-loop embedding provide the inductive bias needed to handle variable timing and shock structure; the remaining requirement is coverage of the target regimes in the training distribution.}

To capture DT interface dynamics, each simulation is initialized at time $t=0$ with a Lagrangian tracer/particle placed at the DT interface, allowing the interface radius $R_i(t)$ and velocity $V_i(t)$ (among other state variables) to be captured along dense output times over the entire time horizon $[0, t_f]$. In our simulations, there are $N_t=10^3$ uniform output times $\mathbf t=\{t_m\}_{m=1}^{N_t}$, where $t_f= 9.9252\, [ns]$. Given that we are interested in the implosion phase of the capsule, for each simulation, we post-process the DT interface trajectory in the following manner: (a) we locate time $t_{m^*}:=\arg\min_{t_m} R_i(t_m)$, i.e., when the radius reaches its minimum before the velocity changes sign, and (b) pad the radius with its most recent value and velocity with zeros, i.e., $R_i(t_m)= R_i(t_{m^*})$ and $V_i(t_m)=0$ for all $m > m^*$. This is done so all drive and interface sequences have the same length, which is required by the proposed NN architectures. Note, if the various interface sequences of variable length were normalized so that $t_{m^*} = t_f$ for each simulation, then $t_{m^*}$ would lose it's physical interpretation, which we avoid by padding. 

We denote our LF dataset by $\D_{\text{LF}} := \{(\mathbf T_{r,n}\,, \mathbf R_{i,n}^{\text{LF}}, \mathbf V_{i,n}^{\text{LF}})\}_{n=1}^{N_{\text{LF}}}$, consisting of $N_{\text{LF}}$ statistically independent (discretely sampled) drives $\mathbf T_{r,n}$ corresponding DT interface radius $\mathbf R_{i,n}^{\text{LF}}$ and velocity $\mathbf V_{i,n}^{\text{LF}}$ trajectories. Similarly, we have a set of HF data $\D_{\text{HF}} := \{(\mathbf T_{r,n}\,, \mathbf R_{i,n}^{\text{HF}}, \mathbf V_{i,n}^{\text{HF}})\}_{n=1}^{N_{\text{HF}}}$. We are interested in the case when $N_{\text{HF}} \ll N_{\text{LF}}$. Moreover, we restrict the sets of LF and HF drives to be mutually exclusive. From a MF learning perspective, this is the more challenging setting since no LF simulation data, i.e., from $\D_{\text{LF}}$, can be used directly as input to train a HF surrogate. Instead, only LF surrogate predictions are used in training a HF surrogate, which decreases the possibility of bias and/or noise being introduced from LF prediction errors and limited HF training data.

When considering our forward and inverse problems, we consider independent data sets to avoid data leakage. That is, in solving the forward problem, i.e., building the MF surrogate, we randomly partition the data into mutually exclusive low- and high-fidelity training, validation, and test sets $\D_{\text{LF}}^\text{For}=\D_{\text{LF}}^{\text{For, tr}}\cup\D_{\text{LF}}^{\text{For, val}}\cup\D_{\text{LF}}^{\text{For, test}}$ and $\D_{\text{HF}}^\text{For}=\D_{\text{HF}}^{\text{For, tr}}\cup\D_{\text{HF}}^{\text{For, val}}\cup\D_{\text{HF}}^{\text{For, test}}$, where $\D_{\text{LF}}^\text{For}$ is used for the standalone LF surrogate $\F_\text{LF}$, and $\D_{\text{HF}}^\text{For}$, along with the $\F_\text{LF}$ predictions based on HF drives, is used for the map $\F_\text{HF}$, and therefore the full MF surrogate $\F_\text{MF}$. For the various drive estimation inverse problems, we use an independent HF dataset $\D_{\text{HF}}^\text{Inv}=\D_{\text{HF}}^{\text{Inv, tr}}\cup\D_{\text{HF}}^{\text{Inv, val}}\cup\D_{\text{HF}}^{\text{Inv, test}}$, where trajectories generated from $\F_\text{MF}$ predictions are used in place of real HF trajectories during training and validation.

\section{Multi-fidelity reduced-order surrogate framework}
\label{sec:surrogate} 

A dynamic surrogate model is built in three stages around a physically informed ansatz of the material interface of the implosion approximated as an incompressible shell imploding into vacuum (\Cref{subsec:embedding}). Such a model accurately describes the initial \revthree{ballistic} phase of implosion, and can be described by a system of ODEs in radius and velocity. First, this base ODE is augmented with a parameterized controller, which modifies the evolution of the shell interface by modification of the kinetic energy via a forcing function, described in \Cref{subsec:embedding}. For all simulated LF and HF data, we solve for optimized controller coefficients via an ODE optimal control problem to accurately reproduce DT interface dynamics in the rad-hydro simulations with the controlled ODE. Second, we formulate a LF surrogate $\F_\text{LF}$ via a causal NN model that infers LF controller coefficients given a time-dependent temperature drive, resulting in a surrogate LF forward model that maps temperature drive to a 2D ODE that can be rapidly integrated numerically to estimate DT interface dynamics (\Cref{subsec:lf-nn}). Lastly, using a transfer learning approach, a similar architecture is used to build a surrogate $\F_\text{HF}$ to map LF controller predictions to their HF (controller) counterparts via residual learning. Hence, during inference, for a given drive $T_r(t)$, the surrogate $\F_\text{MF}:=\F_\text{HF}\circ\F_\text{LF}$ cheaply predicts a HF controller $\hat P^{\text{HF}}(t)$, which is treated as the source in the base ODE, giving a dynamic prediction for the interface's radius and velocity upon integration.

\subsection{Parameterized embedding}
\label{subsec:embedding}

Consider the mapping of input drives $T_r$ to ICF state variables $\u$. Formally, we have a nonlinear differential operator $\mathcal N: \mathcal T_r \times \mathcal U \to \mathcal Z$ over the triple of Banach spaces $(\mathcal T_r,\mathcal U,\mathcal Z)$, giving rise to a parametric partial differential equation (PDE) of the form $\mathcal N(T_r,\u)=0$ with boundary conditions $\mathcal B(T_r,\u)=0$. Here, $T_r\in\mathcal T_r$ represents the input function, i.e., the temperature drive, and $\u\in\mathcal U$ is the solution to $\mathcal N(T_r,\u)=0$ with prescribed boundary conditions. If there exists a unique solution $\u\equiv \u(T_r)$, then the solution is an operator $G:\mathcal T_r \to \mathcal U$ with $G(T_r) = \u(T_r)$. This is the framework for classical operator learning methods, such as Deep Operator Networks (DeepONets) \cite{Lu2021,Wang2022}, Fourier Neural Operators \cite{DBLP:conf/iclr/LiKALBSA21}, Graph Kernel Networks \cite{DBLP:journals/corr/abs-2003-03485}, and Nonlocal Kernel Networks \cite{You2022}, where a NN architecture is designed to approximate the map $T_r\mapsto \u$. Now, consider an infinitesimal Lagrangian particle of interest $\x_0\in\R^2$ in the PDE's state space, specifically at the DT interface at time $t=0$, and assume the PDE solution $\u$ has induced an embedding in the form of a parameterized nonlinear ODE governing the DT interface $\x(t)$. In other words, we assume the existence of a parametric Banach space $\mathcal P$ and well-defined, appropriately measurable operator $\Pi:\mathcal U\to\mathcal P$ such that, for each $\u\in\mathcal U$ and corresponding parameters $P(t)\in\mathcal P$, $\x_0$ evolves according to 
\begin{align} \label{eq:xode}
    \dot \x(t) = \mathbf h(\x(t),t; P(t)), \qquad \x(0) = \x_0,
\end{align}
for some $\mathbf h: \R^2 \times [0,t_f]\times \mathcal P \to \R^2$ satisfying standard ODE well-posedness assumptions ensuring global existence and uniqueness, e.g., Lipschitz continuity in $\x$, etc. \cite[Ch.~2.2-2.3]{teschl2012}. Here, we assume $P\in\mathcal P$ are Lebesgue measurable and bounded almost everywhere on $[0,t_f]$ so that \cref{eq:xode} may be interpreted in the sense of Carath\'{e}odory [ibid]. Fixing $\x_0$, we denote the ODE solution map by $H_0: \mathcal P \to \mathcal X$, where $\mathcal X$ is the Banach space encoding the particle's trajectories.

Under these assumptions, we seek to approximate the well-defined reduced-order map $E:= H_0 \circ \Pi \circ G: \mathcal T_r \to \mathcal X$ by choosing a fixed ODE model $\mathbf h$ in \cref{eq:xode} based on simplifying approximations of the underlying physics. Thus $H_0$ is fixed and approximating the map $E$ is equivalent to approximating the map $F:=\Pi \circ G: \mathcal T_r \to \mathcal P$. After restricting $\mathcal P$ to be a relatively simple function space (e.g., piecewise constant functions on $[0,t_f]$), for an observed drive and interface trajectory, we generate a corresponding $\tilde P$ which we refer to as a \emph{controller} by solving a trajectory-tracking optimal control problem \cite{Lin2014} constrained by the ODE \cref{eq:xode}. That is, under an ODE model assumption, we determine certain parameters $\tilde P(t)$ such that \cref{eq:xode} reproduces observed DT interface dynamics for a given simulation. This is repeated (in parallel) over an ensemble of statistically independent realizations, generating data pairs $\{T_{r,n}(t),\tilde P_n(t)\}_n$, which are used in training a NN surrogate $\F$ to approximate $F$ in the single-fidelity setting. The choice of ODE model and corresponding optimization is presented in \Cref{subsec:control}.

\subsection{Semi-analytical Model and Control Framework}
\label{subsec:control}

We choose a base ODE model $\mathbf h$ \cref{eq:xode} from physical insight that for shock driven implosions, after the shock has initially propagated through the shell, the proceeding ``early'' dynamics can be well approximated by incompressible flow models. To that end, we build on the reduced-order ODE model for an imploding 1D incompressible shell derived in \cite{Book1987}, which describes the \revthree{ballistic}-phase evolution of the inner and outer interfaces of an imploding shell with specified initial radii and shell velocity, based on conservation of mass and total kinetic energy. Here, we introduce a time-dependent power function as a source term to inject/expel energy into/out of the imploding shell, which we use as a controller to extend the range of applicability of the model beyond \revthree{ballistic} phase.

We denote the shell's inner and outer radii by $R_i(t)$ and $R_o(t)$, respectively, and the inner and outer velocities by $V_i(t)$ and $V_o(t)$, respectively, where the sign convention is positive radii and negative velocities, and the shell is assumed to have uniform density $\rho(R) \equiv \bar{\rho}$.
To extend the model from \cite{Book1987}, we no longer assume constant total kinetic energy, instead formulating a balance law of initial energy $W_0$ plus external energy added/removed $\hat W(t)$. Then, at all times, the total kinetic energy satisfies $W(t) := W_0 +\hat W(t) = W_0 + \int_0^t P(s)\, ds.$
The resulting ODE system derived in \Cref{app:book-derivation} as an extension of the original model is given by
\begin{align} \label{eq:bk-control}
    \dot R_i &= V_i, \notag \\
    \dot V_i &= \tfrac{-W}{4\pi\bar\rho R_i^4} \cdot\left[3 + \tfrac{2R_i}{R_o} + \left(\tfrac{R_i}{R_o}\right)^2\right] + \tfrac{PV_i}{2W},
\end{align}
where the outer radius is implicitly given by $R_o:=\sqrt[3]{R_c^3 + R_i^3}$. Here, conservation of mass provides the constant $R_c^3$ and ensures the velocity relationship $V_o=(R_i/R_o)^2V_i$. 

The power source $P(t)$ will be parameterized as a piecewise constant controller \cite{Lin2014} with $N_k=121$ uniform knots $\boldsymbol\tau=\{\tau_k\}_{k=1}^{N_k}$ and corresponding coefficients $\mathbf p=\{p_k\}_{k=1}^{N_k-1}$ on the interval $[0,t_f]$. Inspired by the causal nature of the underlying objective, i.e., optimizing the controller to reproduce observed dynamics at time $t=s$ should \emph{not} depend on solution or controller states for time $t>s$, we solve successive optimization problems for $p_k$ in the spline parameterization of $\tilde{P}(t)$ over individual knot intervals $(\tau_k,\tau_{k+1}]$. For $k\geq 1$, we consider a reference/true interface trajectory $\x^\text{ref}(t) = (R_i^\text{ref},V_i^\text{ref})$ (i.e., DT interface trajectories from either $\mathcal{D}_\text{LF}^\text{For}$ or $\mathcal{D}_\text{HF}^\text{For}$), and assume controller coefficients $\{\tilde p_1,\dots,\tilde p_{k-1}\}$ are provided. The subsequent scalar control coefficient $p_k$ is solved by minimizing the ODE-constrained inner velocity integrated squared error (ISE) over $(\tau_k, \tau_{k+1}]$:
\begin{align} \label{eq:bk-wf-control}
    \min_{p_k} J(\x;p_k) := \frac12\int_{\tau_k}^{\tau_{k+1}} \left(V_i(t) - V_i^{\text{ref}}(t)\right)^2 \,dt
\end{align}
subject to \cref{eq:bk-control} with initial condition $\x(\tau_k)=\tilde \x(\tau_k)$. Here, $\tilde\x(\tau_k)$ is the solution to \cref{eq:bk-control} at time $t=\tau_k$ from solving the ODE system over $(0, \tau_k]$ with previously learned controller coefficients $\{\tilde p_1,\dots,\tilde p_{k-1}\}$. We solve the control problem \cref{eq:bk-wf-control} via the adjoint/costate method \cite{Lin2014} in an optimize-then-discretize fashion, which is detailed in \Cref{app:adjoint-control}. 

\begin{remark}
    \revone{The controller parameterization is not sensitive to modest changes in knot count $N_k$: reducing $N_k$ by 5–10 knots yields negligible changes in the optimized solution, whereas larger reductions (e.g., $N_k=61$) introduce approximation error, typically by smoothing sharp transitions during acceleration phases (cf. \Cref{fig:controller}, $6.5$–$7 \,[ns]$). Following standard practice in parameterized control \cite{Lin2014}, $N_k$ can be increased until error falls below a prescribed tolerance; in our case this sweep is inexpensive and trivially parallel. While adaptive knot placement via time-scaling is possible in principle [ibid], \cref{eq:bk-control} is already infinitely stiff (as the radius approaches the axis) but is well handled by implicit integrators. The additional time-scaling needed to optimize knot locations further increases stiffness/ill-conditioning, and we observed non-convergence of the optimization even with implicit integration.}
\end{remark}

\subsection{Low-fidelity Network}
\label{subsec:lf-nn}

Consider the ``high-data" LF regime, where for each simulation we optimize controller coefficients to reproduce the DT interface using a parameterized ODE \cref{eq:bk-control} as discussed in \Cref{subsec:control}. We now target the supervised learning task of using the $T_r(t)$ induced by the laser drive as input to predict $N_k -1$ LF controller coefficients $\mathbf{p}^\text{LF}\in\mathbb{R}^{N_k-1}$ for the piecewise constant controller $P^\text{LF}(t)$. In learning this surrogate $\F_{\text{LF}}$, we consider mutually exclusive LF training, validation, and test sets of sizes $N_{\text{LF}}^{\text{For, tr}}=4\times10^3$, $N_{\text{LF}}^{\text{For, val}}=2\times10^3$, and $N_{\text{LF}}^{\text{For, test}}=2\times10^3$, respectively, where the controller coefficients in these sets have been computed from the corresponding LF interface data via the optimal control approach in \Cref{subsec:control}. 

We adopt a causal sequence-to-sequence encoder-decoder architecture tailored to the physics of the problem: smooth control input, delayed system response, and sharp output transitions. The network consists of three main stages:
\begin{itemize}
  \item \revtwo{\textbf{Temporal convolutional network (TCN).}} A causal (left-padded) 1D convolutional encoder that downsamples and time-aligns the drive with the sequence of controller knots, while extracting diverse temporal features.
  
  \item \revtwo{\textbf{LSTM.}} A multi-layer LSTM that models delayed dynamics and temporal accumulation. \revtwo{An LSTM network is a recurrent neural network (RNN) architecture designed to model temporal/sequential dependencies while mitigating vanishing/exploding gradients via gating mechanisms and internal states. They are inherently causal and learn long-term nonlocal dependencies} \cite{HochreiterSchmidhuber1997}.

  \item \revtwo{\textbf{Multilayer perceptron (MLP)}. An MLP decoder, i.e., a fully connected feedforward network,} that maps latent features to LF controller coefficient predictions $\hat{\mathbf{p}}^\text{LF}$.
\end{itemize}
The convolutional map is a means for causal downsampling, ensuring that only past and current information influences each coarse time step, maintaining physical causality. 

Given that the drive input exhibits local variation in magnitude and timing across samples, we  standardize its scale across the dataset while preserving causal structure and time-local variation by applying z-score normalization independently to each time point \cite{LeCun2012}.
Because controller coefficients contain physically meaningful zeros denoting non-dynamic periods at the beginning and end of each sequence, we instead apply masked global standardization to controller coefficients, where only the dynamic/middle nonzero portion of the coefficients are used to contribute to a mean and standard deviation over samples and knots. This ensures that the standardization is unbiased by sample-to-sample variation in non-dynamic regions \cite{Che2018}. Lastly, we augment both the standardized input and output data with low-level independent and identically distributed zero-mean Gaussian noise, which helps model generalization and training stability over the non-dynamic constant/anchored regions of the data \cite{bishop1995training, goodfellow2016deep}. The noise standard deviation is taken to be $10^{-4}$. Due to low-noise nature of the problem and the need to preserve sharp temporal transitions in the predictions $\hat{\mathbf p}^\text{LF}$, we adopt the Huber loss, which offers robustness to small errors and avoids excessive penalty on outliers, and takes the form
\[
\mathcal{L}_{\text{Huber}}(r) :=
\begin{cases}
\frac{1}{2} r^2 & \text{if } |r| \leq \delta \\
\delta (|r| - \frac{1}{2} \delta) & \text{otherwise},
\end{cases}
\]
where $r$ is the residual between predicted values and true targets. 
\revone{Adopting the Huber loss is based on a number of factors including robustness to outliers (like $L_1$), smoothness and ease of optimization (unlike $L_1$), and better behavior near the optimum. From a learning perspective, the Huber loss scales with the error, whereas under $L_1$, all nonzero residuals share the same gradient magnitude. As a result, the learning dynamics resemble $L_2$ when it is beneficial (i.e., for inliers), while still limiting the influence of large outliers.} We take $\delta=10^{-2}$ based on the distribution of residuals in our validation set. Full details of the architectures and training in \texttt{PyTorch} \cite{pytorch} are provided in \Cref{app:LF}. 

The predicted controller coefficients are post-processed by inputting them into the ODE \cref{eq:bk-control} and numerically integrating to produce the corresponding interface radius and velocity trajectory predictions ($\hat{\mathbf{R}}_i^\text{LF},\hat{\mathbf{V}}_i^\text{LF}$) on the set of $N_t$ discrete times $\mathbf t$.

\subsection{High-fidelity Network}
\label{subsec:hf-nn}

We now consider the supervised learning problem of mapping controller-coefficient predictions for LF dynamics to their corresponding HF counterparts using relatively few HF training samples. Specifically, we take $N_{\text{HF}}^{\text{For, tr}}=3\times10^2$, $N_{\text{HF}}^{\text{For, val}}=10^3$, $N_{\text{HF}}^{\text{For, test}}=10^3$, and remind the reader that these sets are mutually exclusive from the LF data set $\D_{\text{LF}}^\text{For}=\D_{\text{LF}}^{\text{For, tr}}\cup\D_{\text{LF}}^{\text{For, val}}\cup\D_{\text{LF}}^{\text{For, test}}$. This means that the HF network does not receive LF predictions for the same drives that were used in training the LF network, and therefore must fully generalize across different input conditions. This design choice avoids artificially inflated performance from overfitting to ``previously seen" LF controllers. Unlike models that refine LF predictions on shared inputs, this setup reflects a more realistic scenario: a well-trained LF model is deployed as a proxy across a broader input space, and the HF model must infer corrections without access to ground truth LF behavior on its own training set. Moreover, performance in this setting validates the HF surrogate's ability to learn true corrections across the domain, not just memorization of LF failure modes.

The LF predictions are obtained from the pretrained surrogate model $\F_\text{LF}$ that maps temperature drive inputs $\mathbf{T}_r\in\mathbb{R}^{N_t}$ to controller coefficients, which, upon inputting into the ODE \cref{eq:bk-control} and integrating, produce corresponding LF predictions for radius and velocity dynamics of the DT interface. We propose a residual learning architecture for HF controller coefficients based on a 2-layer (vanilla) LSTM and a shallow MLP decoder, which uses a gated residual skip connection from the LSTM output to the final residual prediction, similar to the latter half of the LF surrogate's architecture. This model balances temporal memory and localized expressiveness via the LSTM and MLP \cite[Ch.~6,10]{goodfellow2016deep} while leveraging inductive biases from the known structure of the residuals \cite{Lu2022, Forrester2007, Kennedy2000}. We discuss the modeling choices and architectural trade-offs in detail, and show how the architecture reflects properties of the underlying residual dynamics.

We are given a tuple of HF drives and time-aligned LF and HF controller coefficient sequences:
\[
\left\{ \left(\mathbf{T}_{r,n}\,, \hat{\mathbf{p}}_{n}^\text{LF}, \mathbf{p}_{n}^\text{HF}\right) \right\}_{n\in\mathcal{D}_\text{HF}^\text{For}}, \qquad  \mathbf{T}_r\in\mathbb{R}^{N_t},\quad\hat{\mathbf{p}}^\text{LF},\mathbf{p}^\text{HF} \in \mathbb{R}^{N_k-1},
\]
where each sample corresponds to a drive and its corresponding low- and high-fidelity DT interface dynamics in the imploding ICF capsule. LF predictions $\hat{\mathbf{p}}^\text{LF}$ are outputs of the pretrained LF surrogate $\F_\text{LF}$ from the drive input $\mathbf{T}_r$. 
We aim to learn the mapping $\hat{\mathbf{p}}^\text{LF} \mapsto \mathbf{p}^\text{HF}$. Since $\hat{\mathbf{p}}^\text{LF}$ already approximates $\mathbf{p}^\text{HF}$ well for most of the sequence, we instead learn a model for the residual:
\[
\mathbf{r} := \mathbf{p}^\text{HF} - \hat{\mathbf{p}}^\text{LF}, \quad \text{and predict} \quad \hat{\mathbf{p}}^\text{HF} = \hat{\mathbf{p}}^\text{LF} + \hat{\mathbf r}.
\]
To model the residual $\mathbf r$, we use a causal sequence-to-sequence architecture consisting of:
\begin{itemize}
    \item \revone{\textbf{LSTM.}} A 2-layer LSTM with hidden dimension $N_\ell=128$ to capture long-range temporal dependencies in the residual dynamics.
    \item \revone{\textbf{MLP.}} A shallow MLP decoder: a fully connected network with one hidden layer $(N_\ell \to N_\ell \to 1)$ and a \texttt{ReLU} activation, which is applied to each time step independently.
    \item \revone{\textbf{Residual skip.}} A residual (linear) skip connection from the LSTM output directly to the output, bypassing the MLP, which is gated and initialized ($\alpha\approx0$) in the exact same manner as the MLP skip in \cref{eq:skip}.
\end{itemize}
The residual $\mathbf{r} = \mathbf{p}^\text{HF} - \hat{\mathbf{p}}^\text{LF}$ exhibits the following structure: (a) it is approximately zero at early times, where LF and HF coefficients are aligned, (b) it grows smoothly over time in \revthree{ballistic} or slow acceleration periods of the implosion (see \Cref{fig:controller} upper-right for $t\in[5.5,6.5]\cup[6.75,8]$), reflecting a delayed correction, and (c) it can contain sharp features due to LF and HF misalignment at times of rapid acceleration and deceleration (see $t\approx 5.5,6.5,\text{and }8 \,[ns]$). 
The LSTM is well-suited to capture such non-local temporal dependencies, while the shallow MLP provides expressiveness for local, per-time-step corrections without overfitting the limited training data. The skip connection allows the model to directly use temporal features from the LSTM when the MLP decoder is insufficient or overly smooth. The learnable gate provides a flexible, data-driven mechanism for controlling the contribution of the skip pathway during training. The result is a causal surrogate $\F_\text{HF}$ for the HF controller coefficients, which are post-processed by inputting them into the ODE \cref{eq:bk-control} and numerically integrating to produce the corresponding interface HF radius and velocity trajectory predictions. It is trained with same loss, optimizer, learning rate scheduler, and early stopping criterion used for the LF network described in \Cref{app:LF}, which allowed training to converge in $10^3$ epochs.

\begin{remark}
    Transformer networks have become popular for modeling latent-space dynamics, but in our setting they underperformed LSTMs, likely due to limited HF training data and the  controller sequence length. Transformers are typically better suited to much longer sequences (e.g., thousands of knots) and significantly larger training sets (e.g., thousands of samples).
\end{remark}

\section{Multi-fidelity forward surrogate}
\label{sec:forward} 

Here we demonstrate the accuracy of the forward model pipeline and ML surrogates on the simulated LF and HF data sets. We first demonstrate the accuracy of the ODE embedding and learned controller to reproduce DT interface dynamics from full rad-hydro simulations. We then demonstrate the ability of the LF model to infer controller parameters for the embedding from the temperature drive, and the HF model to accurately correct controller parameters from the LF ML model to reproduce DT interface dynamics for HF data, corresponding to 67 energy groups in the rad-hydro simulations.

\begin{figure}[ht!]
	\centering  
	\begin{subfigure}{.3\textwidth}
  		\centering
  		\includegraphics[width=1\linewidth]{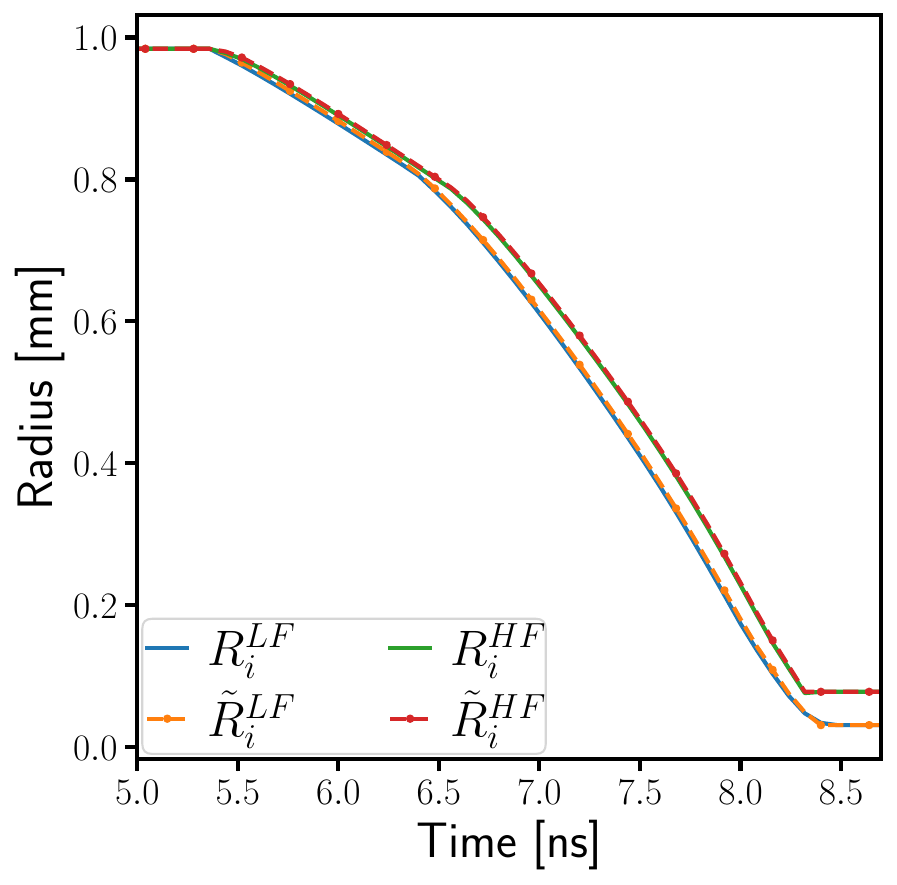}
	\end{subfigure}%
	\begin{subfigure}{.3\textwidth}
 		 \centering
  		\includegraphics[width=1\linewidth]{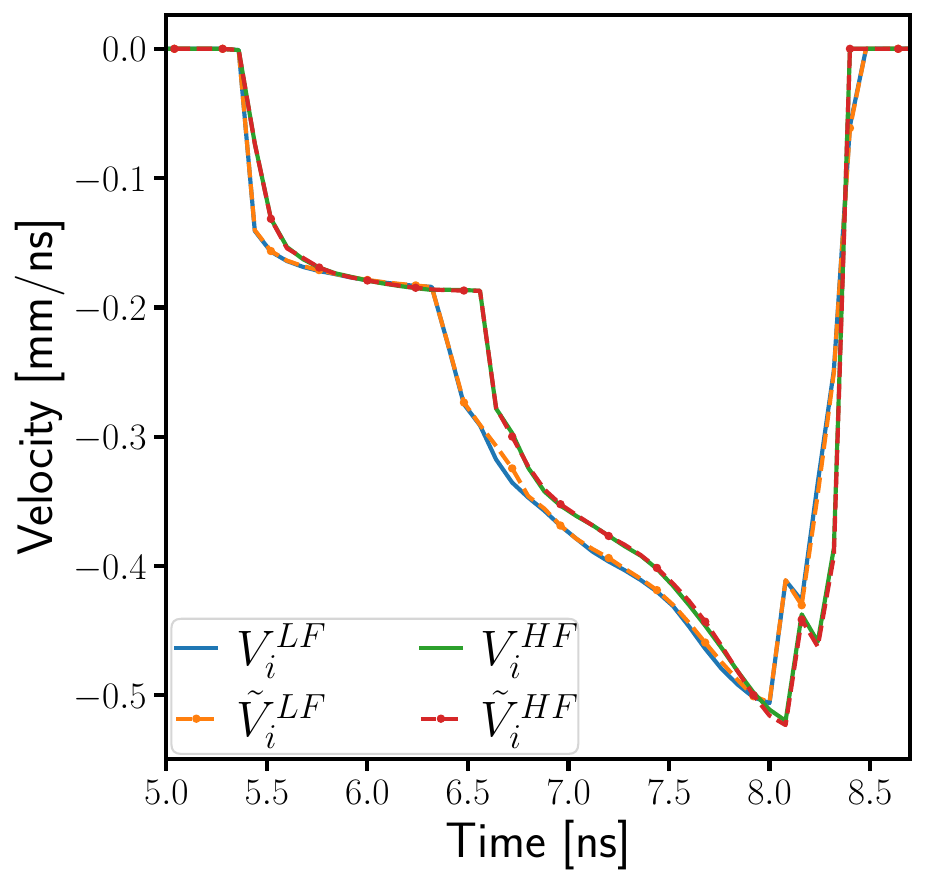}
	\end{subfigure}%
	\begin{subfigure}{.3\textwidth}
 		\centering
  		\includegraphics[width=1\linewidth]{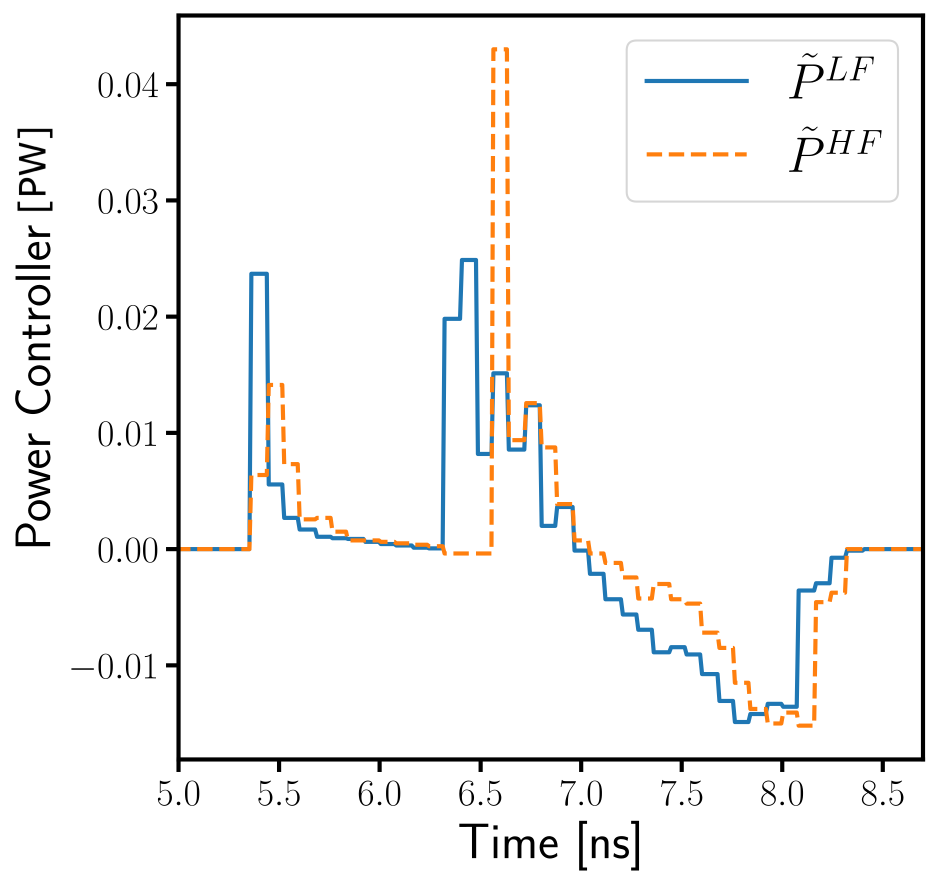}
  	\end{subfigure} \\
    \begin{subfigure}{.3\textwidth}
  		\centering
  		\includegraphics[width=1\linewidth]{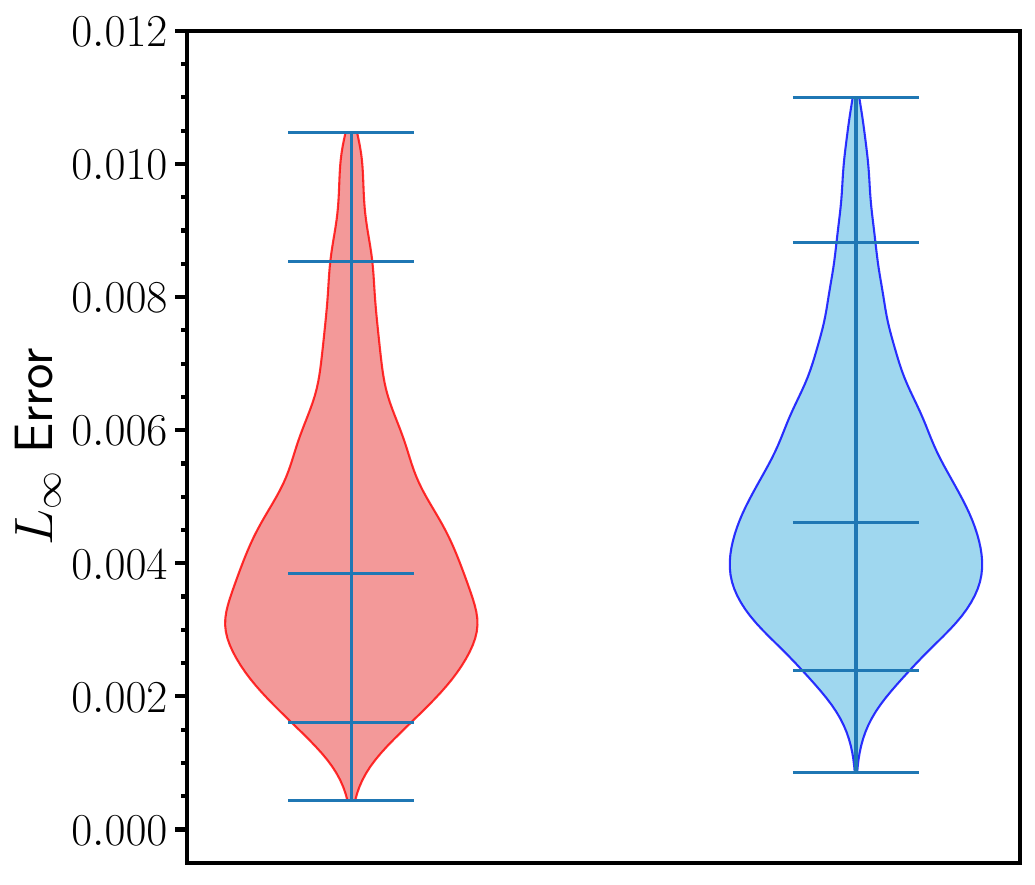}
	\end{subfigure}%
	\begin{subfigure}{.3\textwidth}
 		 \centering
  		\includegraphics[width=1\linewidth]{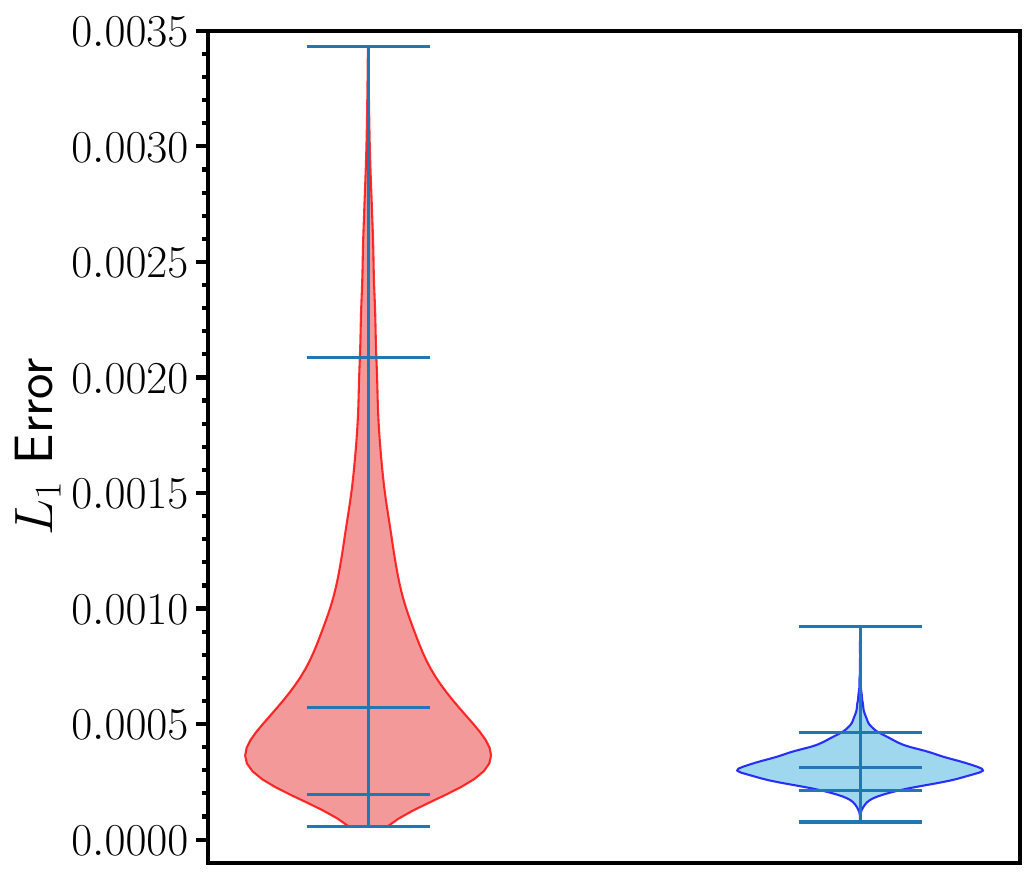}
	\end{subfigure}%
	\begin{subfigure}{.3\textwidth}
 		\centering
  		\includegraphics[width=1\linewidth]{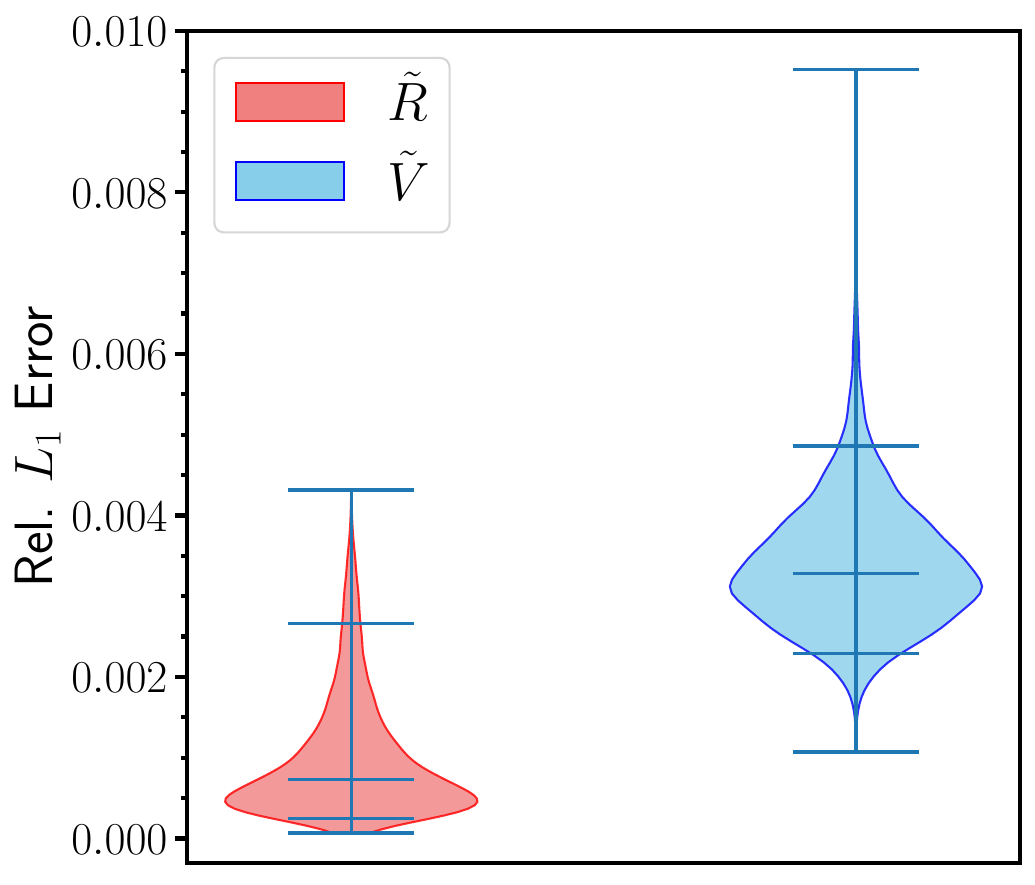}
  	\end{subfigure}
\caption{(Top Row) Reference data versus controlled solution $\tilde\x(t;\tilde{P})=(\tilde{R}_i(t),\tilde{V}_i(t))$ for radius \revtwo{in $[mm]$} (left) and velocity \revtwo{in $[mm/ns]$} (middle) with learned controller $\tilde{P}$ \revtwo{in $[PW]$} (right) corresponding to maximum error in the $L_\infty$ norm, which is $\approx 10 \,[\mu m/ns]$ at $t\approx 6.7 \,[ns]$ in the LF velocity. The corresponding HF controller and solutions are also plotted. (Bottom Row) Controller solution error distributions (with denoted $5^\text{th}$, $50^\text{th}$, and $95^\text{th}$ percentiles) for radius \revtwo{in $[mm]$} (red) and velocity \revtwo{in $[mm/ns]$} (blue) computed via $L_\infty$ (left), $L_1$ (middle), and relative \revtwo{(nondimensional)} $L_1$ metrics. Note, the difference between LF and HF errors are not statistically significant, and therefore their errors have been aggregated.}
\label{fig:controller}
\end{figure}

\Cref{fig:controller} displays the optimal LF and HF controllers $\tilde P$ and the solutions $(\tilde{\mathbf{R}}_i,\tilde{\mathbf{V}}_i)$ to \cref{eq:bk-control} at times $\mathbf{t}$ corresponding to the worst-case $L_\infty$ error over all radius and velocity trajectories from all $(\approx)$ $10^4$ available low- and high-fidelity simulations. This error occurs at $t\approx 6.7 \,[ns]$ in the LF velocity, seen in the top row, middle column. The error distributions for both controlled radius and velocity are also provided (bottom row) on absolute scales via the $L_\infty$ (left) and $L_1$ (middle) norms as well as relative error (right) via the $L_1$ norm (i.e., $||\mathbf{R}_{i,n}-\tilde{\mathbf{R}}_{i,n}||_\infty$, $||\mathbf{R}_{i,n}-\tilde{\mathbf{R}}_{i,n}||_1/N_t$, and $||\mathbf{R}_{i,n}-\tilde{\mathbf{R}}_{i,n}||_1/||\mathbf{R}_{i,n}||_1$, respectively), for radius and likewise for velocity, where the norms are taken over the temporal dimension for each sample $n\in\D_\text{LF}^\text{For}\cup\D_\text{HF}^\text{For}$. We note that the median $L_\infty$ error is less than $4 \,[\mu m]$ for radius and $5 \,[\mu m/ns]$ for velocity, while the maximum $L_\infty$ error is approximately $10.5 \,[\mu m]$ and $11 \,[\mu m/ns]$ for radius and velocity, respectively. These are well below the resolution available in practice, highlighting the accuracy of the controller approach for tracking observed DT interface dynamics. The median and maximum relative errors for radius are less than $0.1\%$ and $0.5\%$, respectively, while approximately $0.3\%$ and $1\%$ for velocity.

\begin{remark} \label{rmk:plots}
    \revtwo{There is a significant delay between the time the drive starts injecting energy into the outer part of the NIF capsule (i.e., $t=0$) and the time at which the DT interface actually begins to move inwardly. Hence, the interface radius remains constant and the velocity zero for the first few nanoseconds, e.g., between $3.5$ and $6 \,[ns]$ for our data. Therefore, in our figures displaying radius and velocity trajectories, we omit these initial constant periods to emphasize regimes with dynamic behavior in radius and velocity.} 
    
    Additionally, to accurately describe the implosion physics, it is critical to capture low-regularity regions in the radius and velocity evolution, which is best measured in an $L_\infty$ sense. We have also included $L_1$ and relative $L_1$ error (i.e., mean absolute error (MAE)) in \Cref{fig:controller} as an example of a (global) temporally aggregated metric; moving forward we only consider relative $L_1$ (i.e., relative MAE). $L_1$ is better suited for our problems than $L_2$ (e.g., relative mean square error (MSE) or root MSE) since the interface dynamics (both radius and velocity) have skewed distributions. Moreover, relative $L_1$ is less sensitive to spikes and is more stable in low-noise regimes. Note, this does not contradict the use of the $L_2$ metric in the control formulation \cref{eq:bk-wf-control} since the ISE there is minimized over single knot intervals rather than the entire time horizon $(0,t_f]$.
\end{remark}

\begin{figure}[ht!]
	\centering  
	\begin{subfigure}{.3\textwidth}
  		\centering
  		\includegraphics[width=1\linewidth]{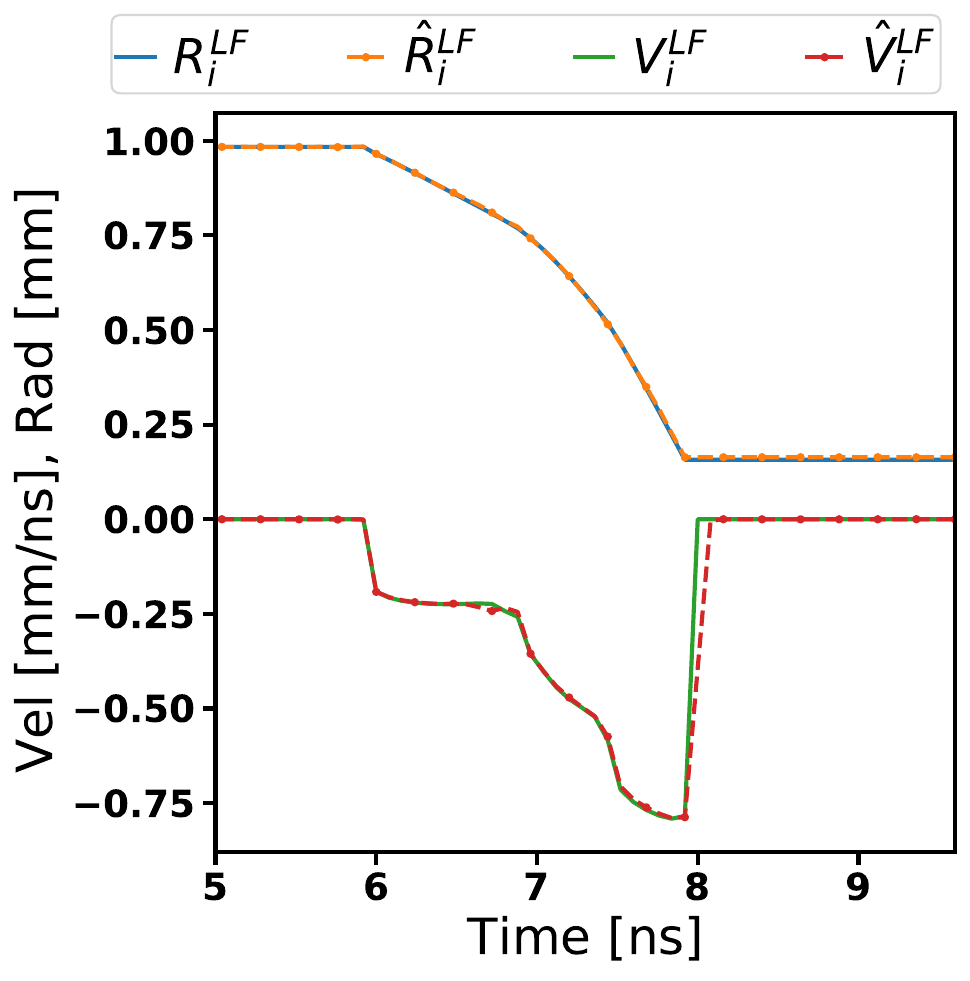}
	\end{subfigure}%
	\begin{subfigure}{.3\textwidth}
 		 \centering
  		\includegraphics[width=1\linewidth]{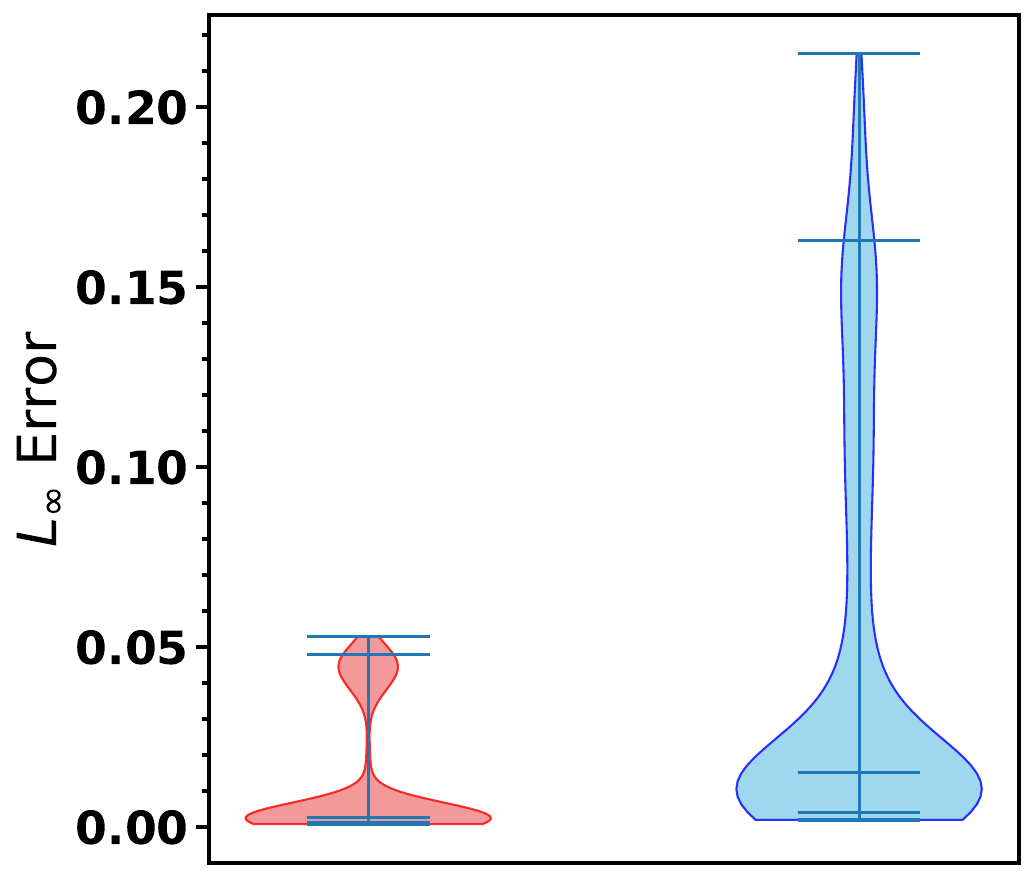}
	\end{subfigure}%
	\begin{subfigure}{.3\textwidth}
 		\centering
  		\includegraphics[width=1\linewidth]{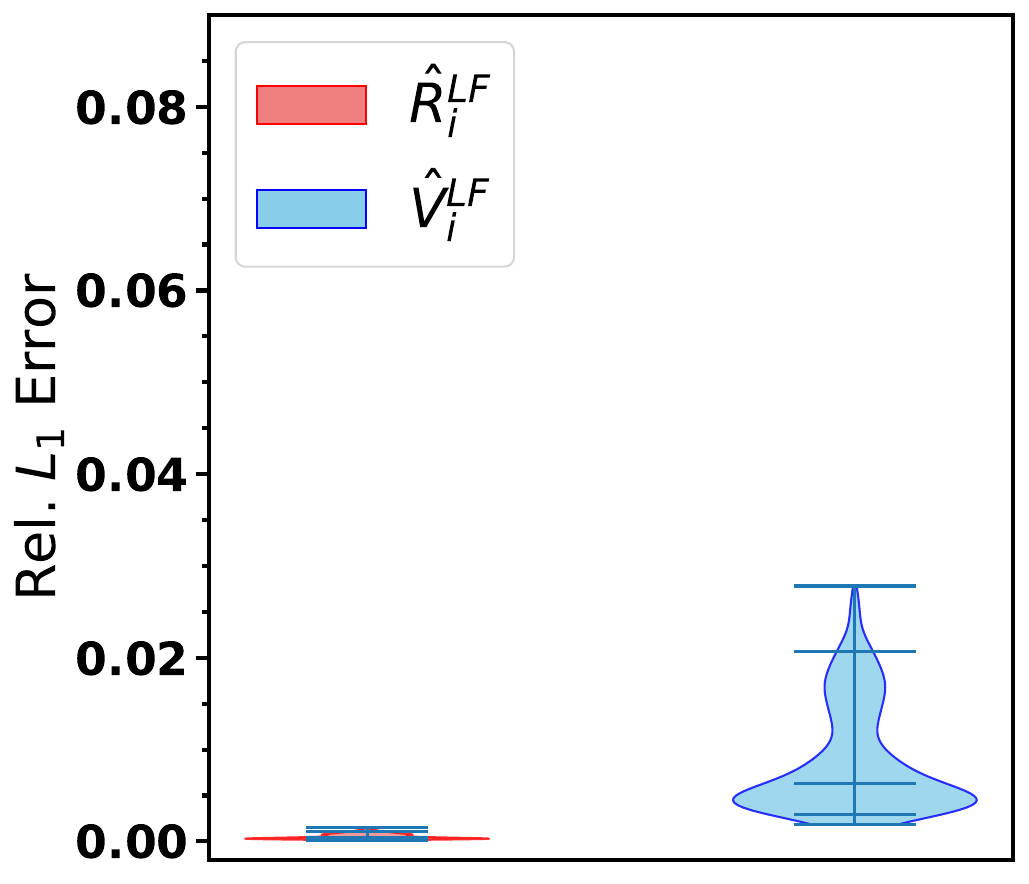}
  	\end{subfigure}
\caption{(Left) Worst-case LF test-set predictions (in $L_\infty$) for radius $\hat{\mathbf{R}}_i^{\text{LF}}$ and velocity $\hat{\mathbf{V}}_i^{\text{LF}}$, which come from two different test samples. They are computed by numerically integrating \cref{eq:bk-control}, using $\F_\text{LF}$ predictions $\hat{\mathbf p}^\text{LF}$ for the controller coefficients $\mathbf p$. Test error distributions for radius \revtwo{in $[mm]$} (red) and velocity \revtwo{in $[mm/ns]$} (blue), with denoted $5^\text{th}$, $50^\text{th}$, and $95^\text{th}$ percentiles, computed via $L_\infty$ (middle) and relative \revtwo{(nondimensional)} $L_1$ (right) metrics.}
\label{fig:LF_error}
\end{figure}

\Cref{fig:LF_error} displays the worst-case test-set radius and velocity profiles integrated from LF controller coefficient predictions in the $L_\infty$ norm, which occur late-time for both samples. The maximum error for the worst-case radius is approximately $50\,[\mu m]$ and occurs at $t\approx 9.3\, [ns]$, while the worst-case velocity error is approximately $0.22\,[mm/ns]$, or $220\,[\mu m/ns]$, at $t\approx 8\,[ns]$. Both cases correspond to under-sampled tails of the training distribution, where the peak temperature drives have very large peaks at over $350\,[eV]$. This causes near immediate deceleration and transition into the outgoing phase, which can be seen via the velocity plot at $t\approx 8\,[ns]$. These transitions are difficult for the surrogate to capture when only a few such cases appear in training. \revone{Moreover, LSTMs are known to smooth sharp, near-discontinuous transitions \cite[Ch.~10]{goodfellow2016deep}; the appended MLP helps preserve these features, as described in \Cref{app:LF}.} Overall, the LF surrogate is quite accurate as seen by the test error distributions. They do exhibit right-skewed tails; however, this is expected for a network that was not optimized for rarer cases. The bulk of the distribution however is excellent with median errors of $2\,[\mu m]$ and $0.04\%$ for the radius and $15\,[\mu m/ns]$ and $0.6\%$ for velocity in $L_\infty$ and relative $L_1$ metrics, respectively. Moreover, the distribution of the test residuals (flattened over samples and time) is roughly Gaussian with mean $-2\times10^{-5}$ and standard deviation $5.6\times10^{-3}$, indicating little to no predictive bias. Additionally, the sample autocorrelation function (ACF) for the residuals (averaged over samples) lies within the $95\%$ confidence interval for Gaussian white noise for all lags, providing evidence that (average) prediction errors are uncorrelated in time. Moreover, the variance-weighted coefficient of determination $R^2$ over the test set is $0.999$ and $0.99$ for the radius and velocity, respectively. Such accuracy is quite sufficient for training the (simpler) HF surrogate with limited data.

\begin{figure}[ht!]
	\centering  
	\begin{subfigure}{.3\textwidth}
  		\centering
  		\includegraphics[width=1\linewidth]{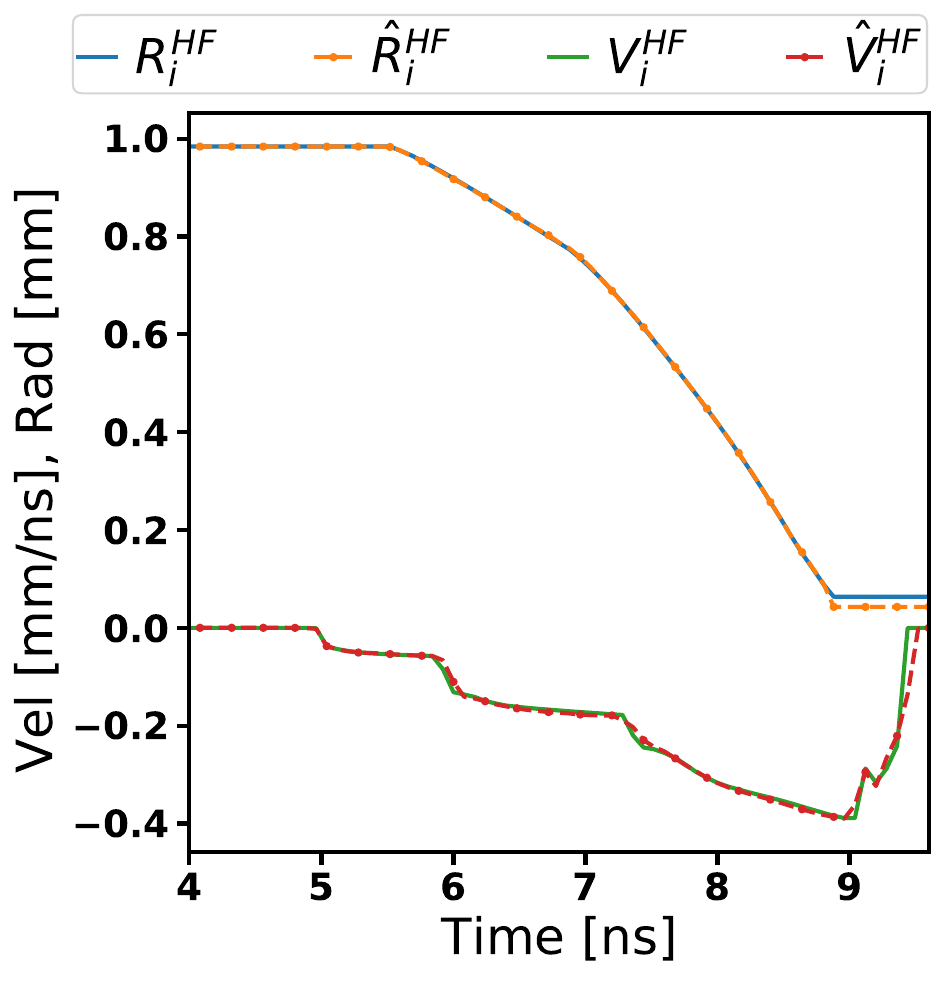}
	\end{subfigure}%
	\begin{subfigure}{.3\textwidth}
 		 \centering
  		\includegraphics[width=1\linewidth]{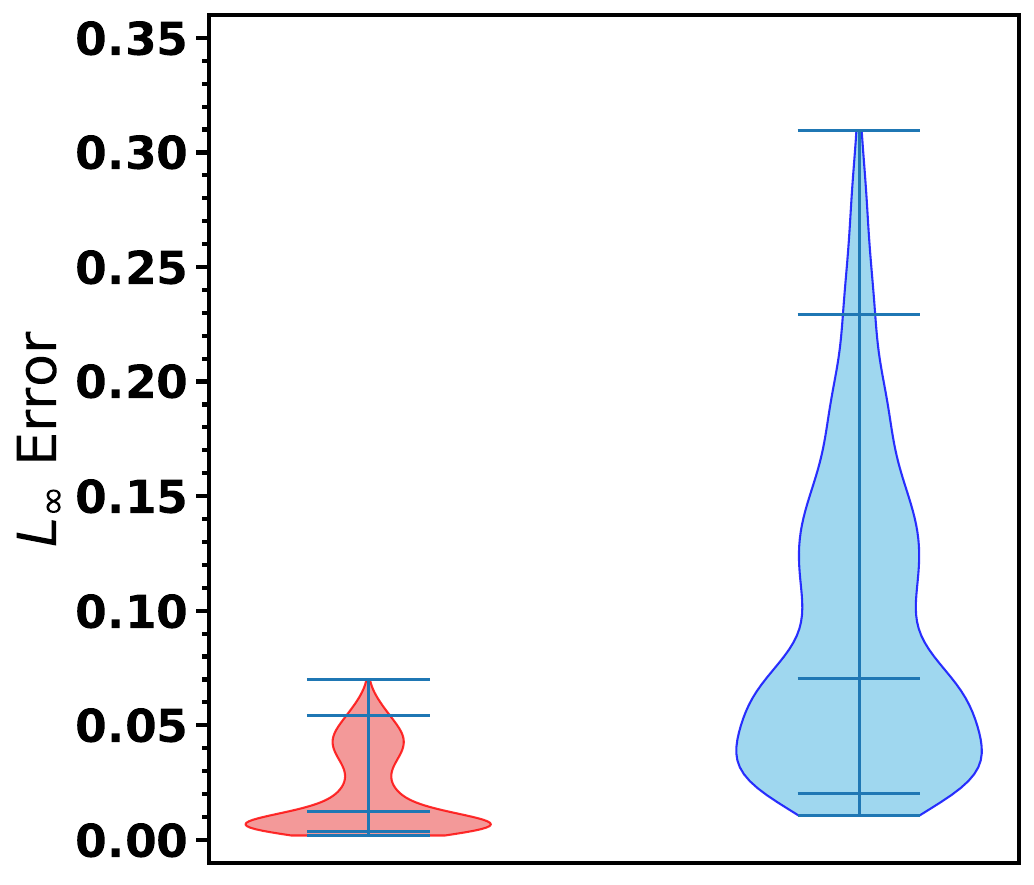}
	\end{subfigure}%
	\begin{subfigure}{.3\textwidth}
 		\centering
  		\includegraphics[width=1\linewidth]{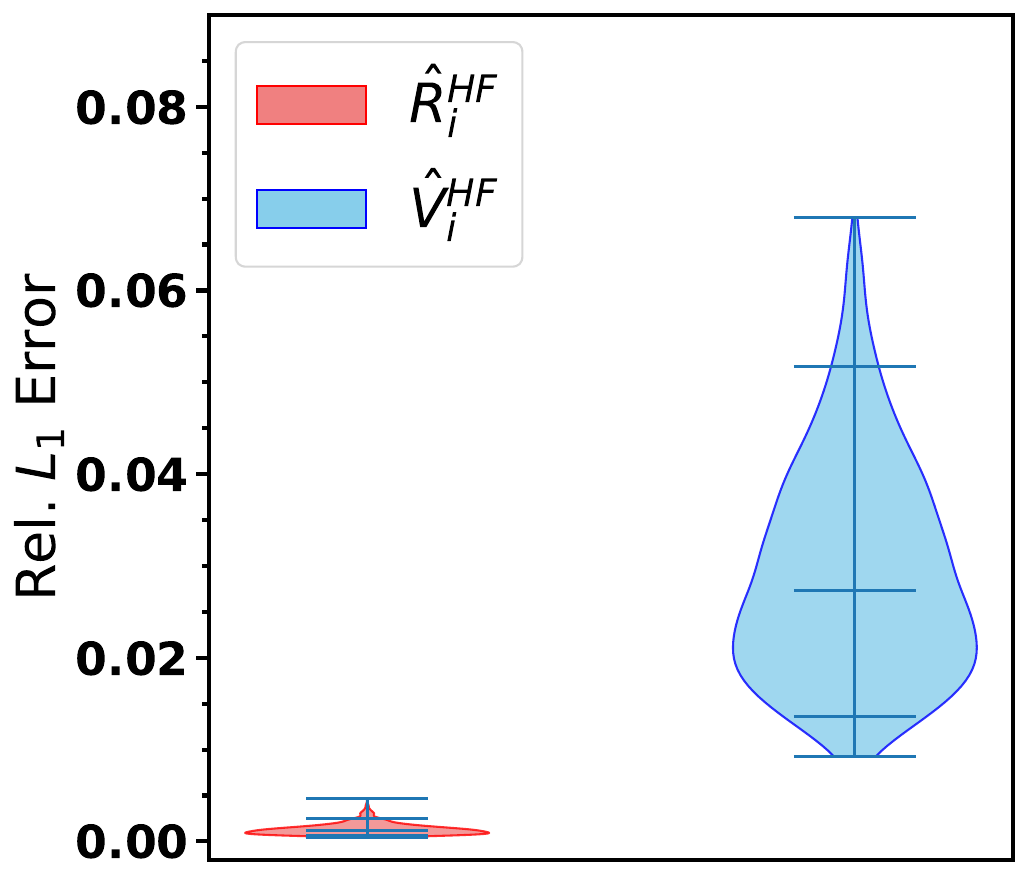}
  	\end{subfigure}
\caption{(Left) Worst-case HF test-set predictions (in $L_\infty$) for radius $\hat R_i^{\text{HF}}$ and velocity $\hat V_i^{\text{HF}}$, which come from two different test samples. They are computed by numerically integrating \cref{eq:bk-control} using $\F_\text{HF}$ predictions for $\hat{\mathbf p}^\text{HF}$ the controller coefficients $\mathbf p$. Test error distributions for radius \revtwo{in $[mm]$} (red) and velocity \revtwo{in $[mm/ns]$} (blue), with denoted $5^\text{th}$, $50^\text{th}$, and $95^\text{th}$ percentiles, computed via $L_\infty$ (middle) and relative \revtwo{(nondimensional)} $L_1$ (right) metrics.}
\label{fig:HF_error}
\end{figure}

\Cref{fig:HF_error} displays the worst-case test-set radius and velocity profiles integrated from HF controller coefficient predictions in the $L_\infty$ norm, which occur at the end of the implosion phase for both samples. The maximum error is just under $70\,[\mu m]$ for radius and is approximately $0.31\,[mm/ns]$, or $310\,[\mu m/ns]$, for velocity, both occurring during the very sharp deceleration periods. While not as accurate as the LF surrogate, the HF surrogate performs remarkably well given the size of its training set. The test error distributions are provided in $L_\infty$ (middle) and relative $L_1$ (right) metrics. The median errors are $13\,[\mu m]$ and $0.1\%$ for the radius and $73\,[\mu m/ns]$ and $2.7\%$ for velocity in $L_\infty$ and relative $L_1$ metrics, respectively. Lastly, the variance-weighted $R^2$ coefficients over the test set are $0.997$ and $0.96$ for the radius and velocity, respectively. Previous statements regarding the distribution and average sample ACF of the LF residuals also hold for the HF residuals, with the only difference being the mean and standard deviation in the HF case are approximately one order of magnitude larger than in the LF case.

\section{Inverse problems: drive estimation}
\label{sec:param-nn}

We consider supervised learning problems for estimating time-dependent temperature drives $\mathbf{T}_r\in\mathbb{R}^{N_t}$ from HF DT interface trajectories, where the reduced-order ODE \cref{eq:bk-control} together with the causal MF surrogate $\F_{\text{MF}}:=\F_{\text{HF}}\circ \F_{\text{LF}}$ for controller coefficients are used to generate a HF training set (of radii and/or velocities) sufficiently large to provide high-accuracy estimation. After independent z-score standardization at each time, the drives are innately low-dimensional, with $99\%$ of explained variance being explained by the first $N_d=4$ principal components. Hence, to simplify the learning frameworks in all problems that follow, we focus on predicting these 4 principal components of the drive $\mathbf{T}_r^\text{PCA}\in\mathbb{R}^{N_d}$, where the inverse PCA and standardization are applied {\emph{ex post facto}} to obtain the estimated drives in their original domain (i.e., $\hat{\mathbf{T}}_r$ at discrete times $\mathbf t$) for test error evaluation.

We consider two main problems. First, in \Cref{subsec:dense-nn}, we consider (PCA) drive estimation from complete HF DT interface trajectories. However, in experimental settings, only a small number of temporal snapshots of the interface are typically available to infer a temperature drive. Thus, in \Cref{subsec:sparse-nn}, we learn a proxy for optimal temporal snapshots from which to infer a drive, i.e., soft ``top K" selection, and then infer the (PCA) drive from HF interfaces at the selected times. Within the second problem we consider the two subproblems of using only radius versus using both radius and velocity to perform drive estimation. This is also motivated by the experimental setting, where only the interface radius is typically available, but there is interest in knowing if velocity could improve drive estimation, and, if so, by how much.

In solving the three inverse problems, we consider a data set $\D_{\text{HF}}^\text{Inv}=\D_{\text{HF}}^{\text{Inv, tr}}\cup\D_{\text{HF}}^{\text{Inv, val}}\cup\D_{\text{HF}}^{\text{Inv, test}}$, independent of the data sets $\D_{\text{LF}}^\text{For}$ and $\D_{\text{HF}}^\text{For}$ used in training, validating, and testing the LF and HF modules of the MF surrogate $\F_\text{MF}$. All inverse problems use the same training, validation, and test sets of sizes $N_{\text{HF}}^{\text{Inv, tr}}=4\times10^3$, $N_{\text{HF}}^{\text{Inv, val}}=10^3$, and $N_{\text{HF}}^{\text{Inv, test}}=10^3$, respectively. We note that radius and/or velocity inputs for all inverse problems are standardized (separately) in precisely the same manner as the controller coefficients in the LF and HF modules of the forward model, i.e., via masked global standardization over the dynamic periods, described in \Cref{subsec:lf-nn}. Additionally, due to limited HF data, only the test set $\D_{\text{HF}}^{\text{Inv, test}}$ contains true radius and velocity trajectories of the DT interface. While the various inverse models are trained and validated with input data from surrogate forward model $\F_{\text{MF}}(\mathbf T_r)$ predictions for HF radius and/or velocity trajectories, the presented errors correspond to true HF trajectories being used as input during test/inference mode. This mimics the real-world scenario of wanting to estimate the drive having observed a HF interface trajectory. Moreover, using true HF dynamics rather than forward model predictions prevents bias from the forward model leaking into the test set (a.k.a, the ``inverse crime"), which would otherwise give falsely optimistic test errors.

\subsection{Optimizing drive -- dense-time networks}
\label{subsec:dense-nn}

In our first inverse problem, we estimate drive principal components directly from entire HF interface trajectories, i.e., from both radius and velocity at all $N_t$ discrete times in $\mathbf t$, via a sequence-to-static encoder–decoder network $\mathcal{I}_\text{D}$. The network consists of three main stages:
\begin{itemize}
  \item \revone{\textbf{LSTM.}} A 3-layer LSTM encoder with hidden dimension $N_\ell=32$.
  \item \revone{\textbf{Attention pooling.} A \texttt{SoftMax} pooling mechanism and a shallow nonlinear bottleneck to dynamically select relevant latent features for static regression.}
  \item \revone{\textbf{MLP.}} A deep MLP decoder with a learnable residual skip connection.
\end{itemize}
where our design rationale stems from the need to capture long‐range temporal patterns while selectively focusing on salient time steps. The network is detailed in \Cref{app:inv1} and trained with same optimizer, learning rate scheduler, and early stopping criterion that was used for the LF and HF modules of the surrogate, which allowed training to converge in approximately $1.4\times10^3$ epochs using the MSE loss. 

\begin{figure}[h!]
	\centering  
	\begin{subfigure}{.35\textwidth}
  		\centering
  		\includegraphics[width=1\linewidth]{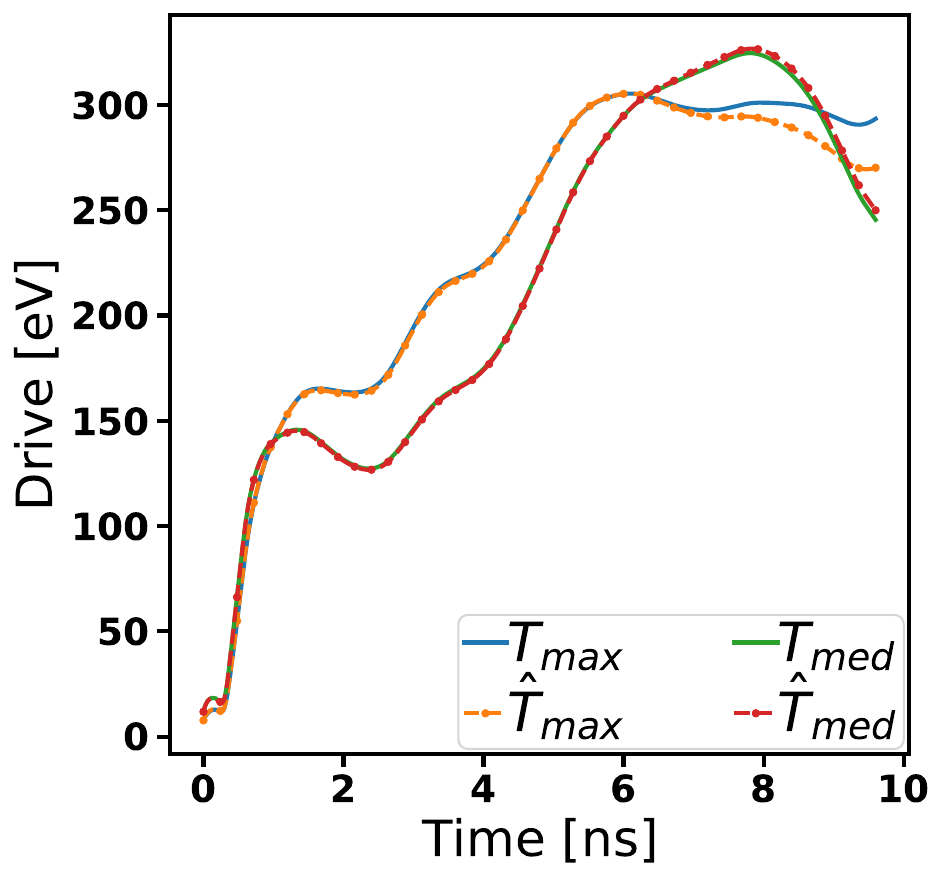}
	\end{subfigure}%
	\begin{subfigure}{.38\textwidth}
 		 \centering
  		\includegraphics[width=1\linewidth]{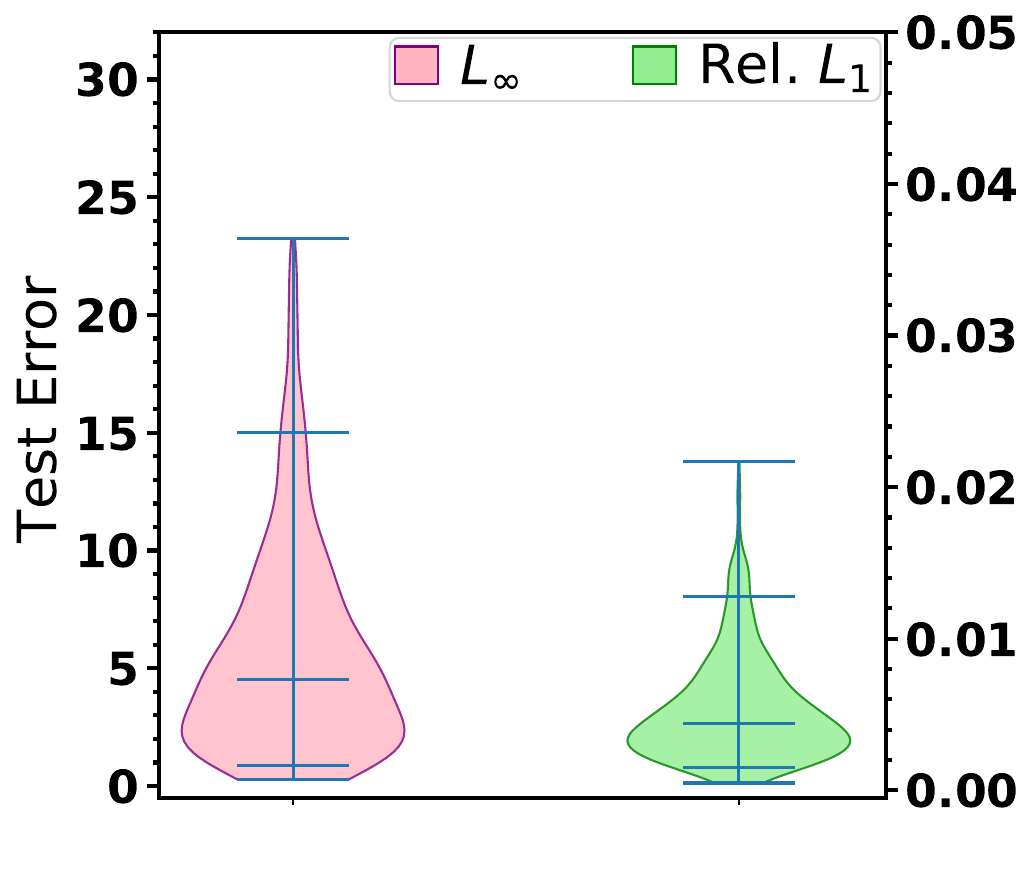}
	\end{subfigure}%
\caption{(Left) Median- and worst-case drive test-set predictions (in $L_\infty$) for inverse model $\mathcal{I}_\text{D}$. (Middle)  Test error distributions via $L_\infty$ \revtwo{in $[eV]$} (pink) and relative $L_1$ \revtwo{(nondimensional)} (green) metrics (with denoted $5^\text{th}$, $50^\text{th}$, and $95^\text{th}$ percentiles).}
\label{fig:dr_error1}
\end{figure}

\Cref{fig:dr_error1} displays the median- and worst-case test-set predictions for the drive in the $L_\infty$ norm, which both occur at $t = t_f = 9.6\,[ns]$, i.e., the final time. The $L_\infty$ error is approximately $23.2\,[eV]$ for the worst case and $4.5\,[eV]$ in the median case. The full $L_\infty$ test-error distribution is provided (pink) as well as its counterpart in the relative $L_1$ metric (green). The median and maximum relative $L_1$ errors are $0.4\%$ and $2.2\%$, respectively. Additionally, the variance-weighted $R^2$ coefficient over the test set is $0.98$. 

\revtwo{Maximal estimation error is expected at late times due to (a) accumulation of small forward-model errors and (b) the causal delay between changes in the drive and their observable effect on interface dynamics. As the drive time approaches the end of the padded/implosion-only interface trajectory, there is insufficient downstream radius and velocity information to constrain it, so late-time drive recovery is inherently unreliable. Moreover, with a causal forward model and an observation limited to $t\le s$, one cannot expect to recover the drive beyond the latest time whose effects are visible within the observation window (roughly $T-\Delta$, where $\Delta$ is the effective response delay). Consequently, inverse-drive uncertainty (and typically error) grows toward the end of the inferred drive and peaks at the final time.} 

\begin{figure}[h!]
	\centering  
  	\includegraphics[width=.32\linewidth]{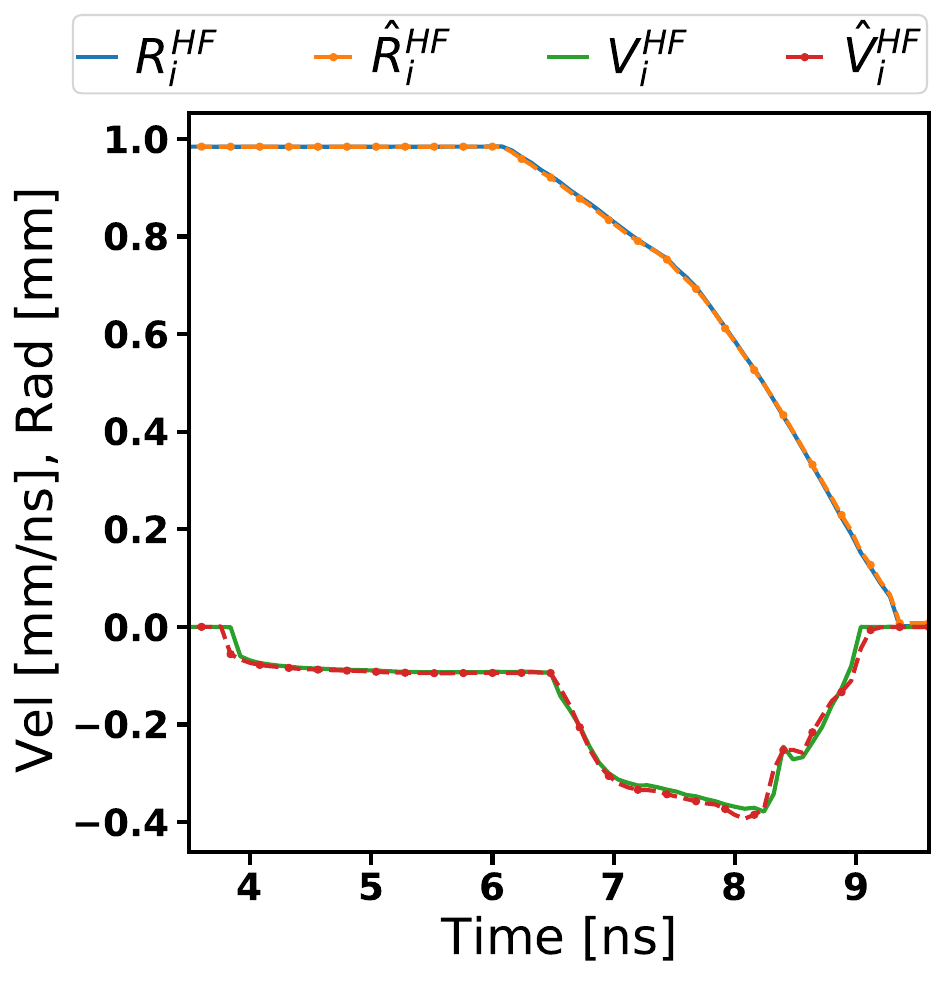}
  	\includegraphics[width=.32\linewidth]{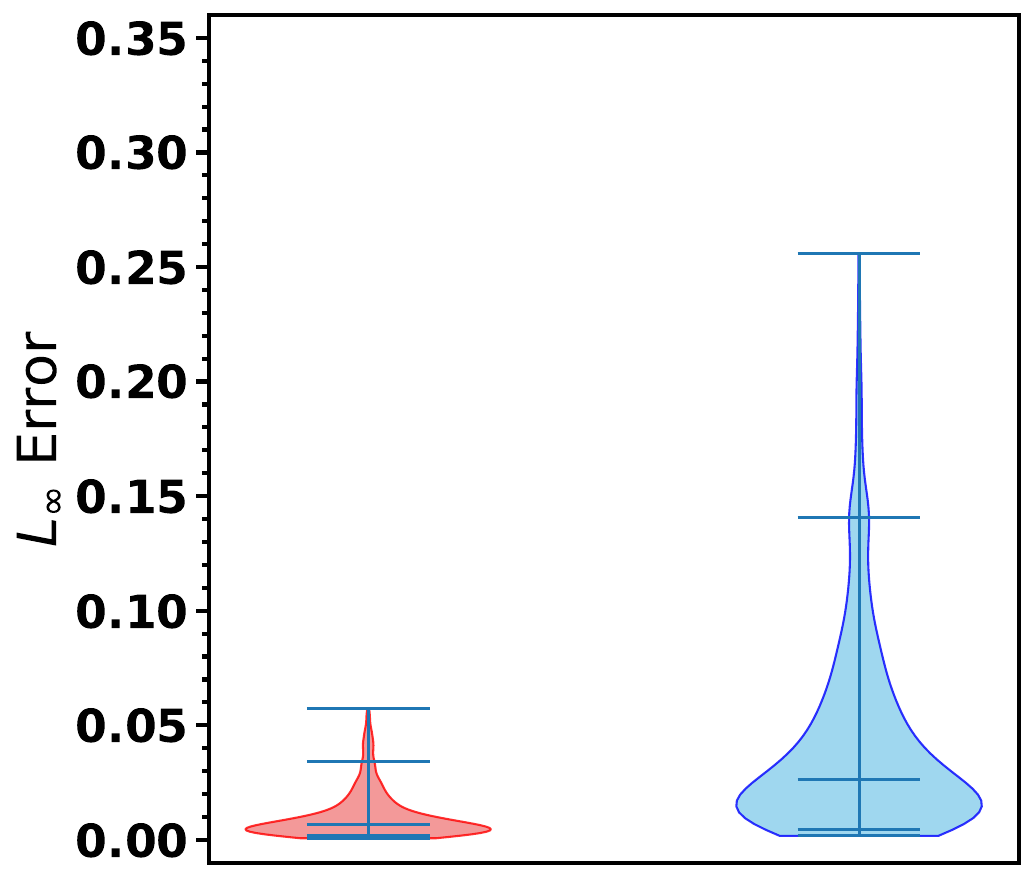}
  	\includegraphics[width=.32\linewidth]{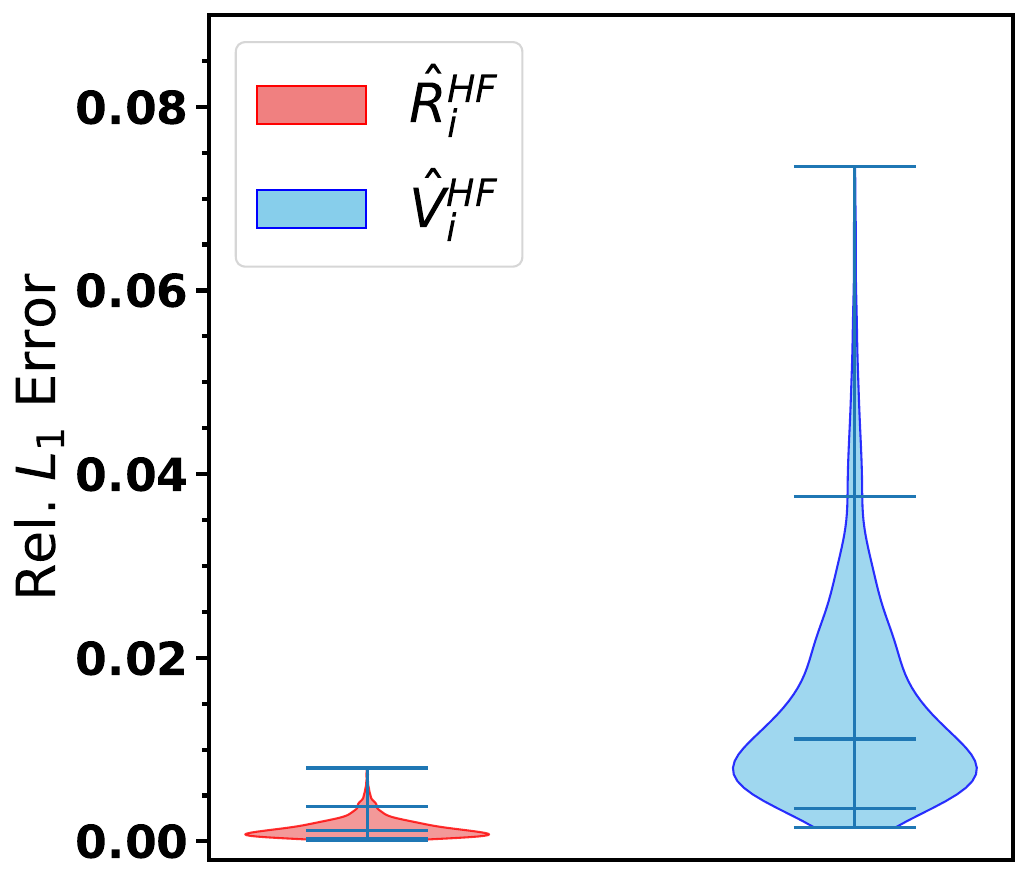}
    \\
  	\includegraphics[width=.32\linewidth]{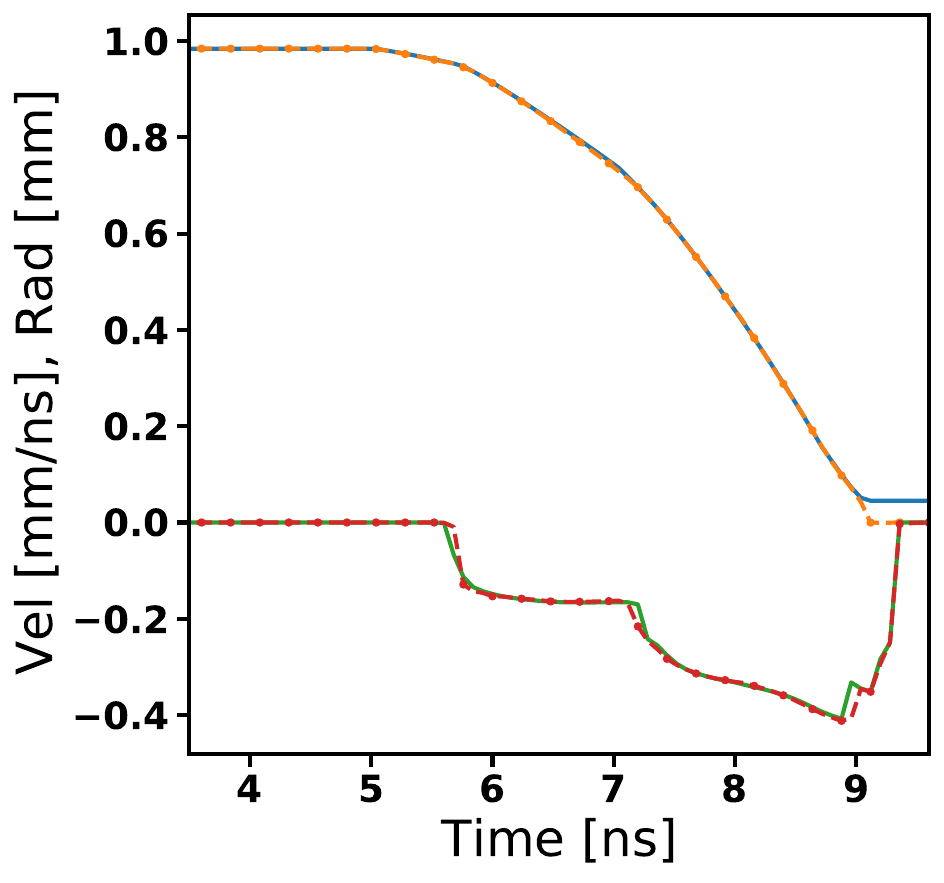}
  	\includegraphics[width=.32\linewidth]{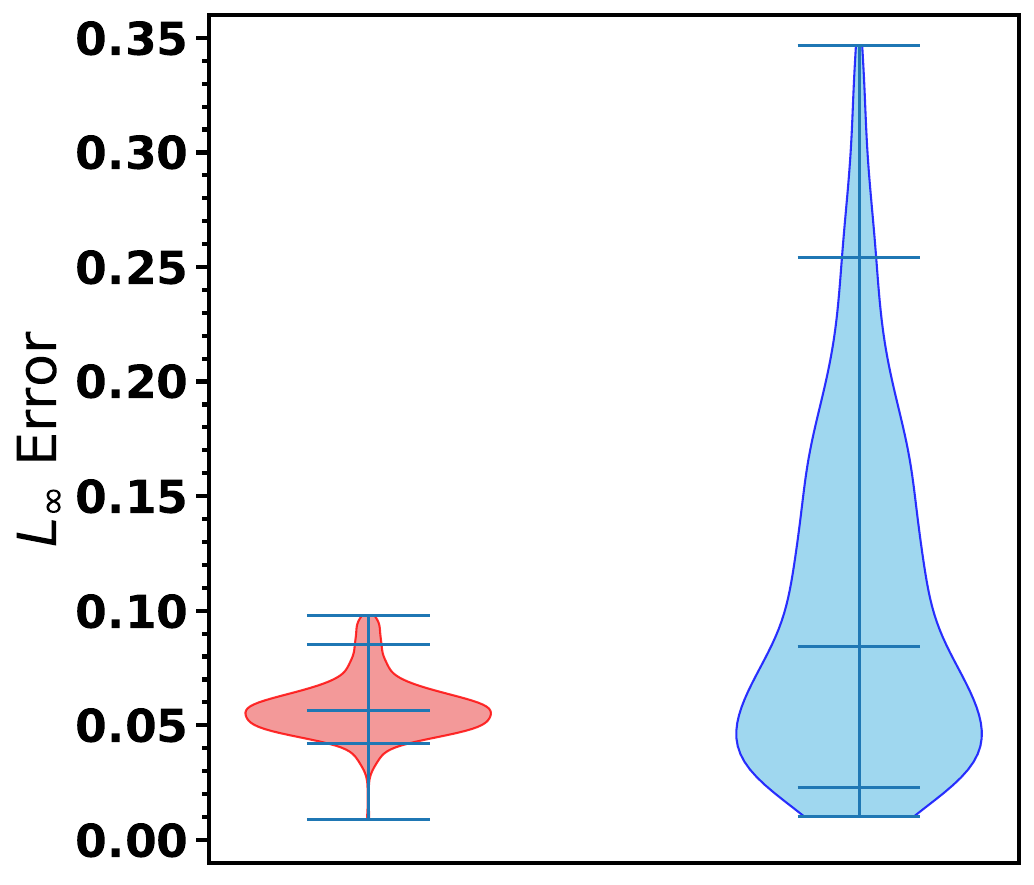}
  	\includegraphics[width=.32\linewidth]{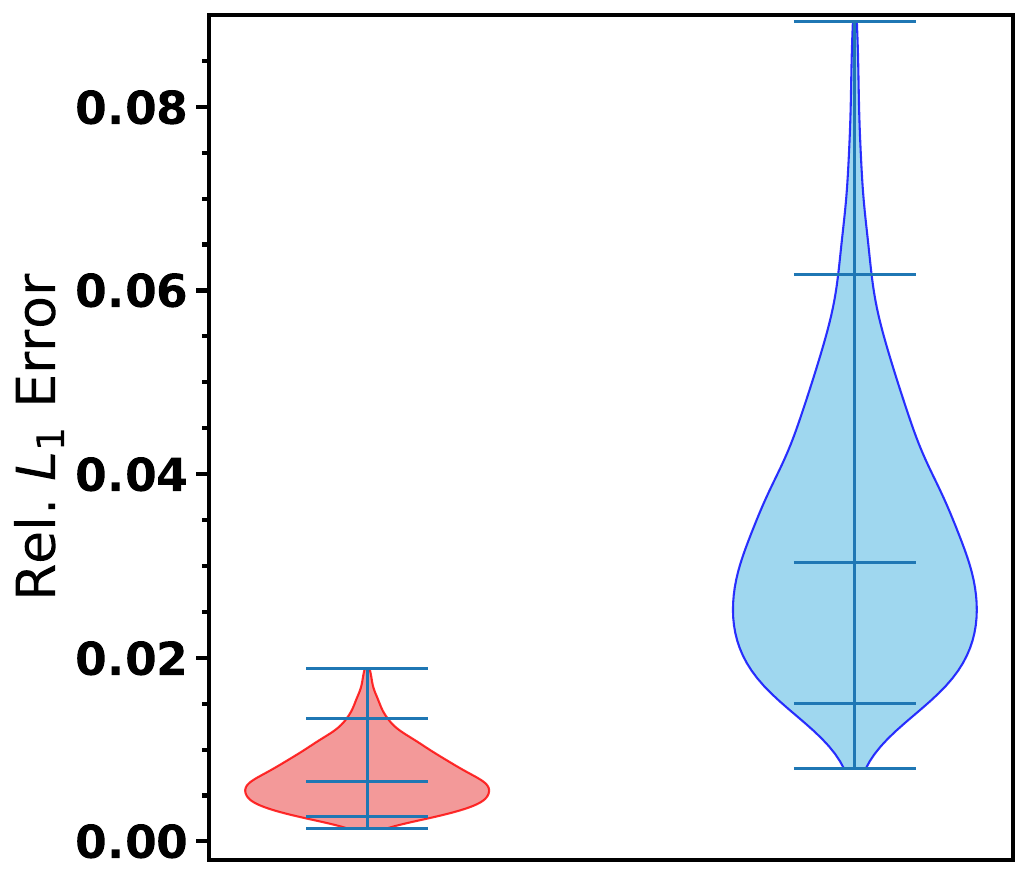}
\caption{Cycle-consistency errors for the forward model via estimated test-drives $\hat{\mathbf{T}}_r$ from inverse model $\mathcal I_\text{D}$. (Top Row) $\F_\text{MF}$'s sensitivity/stability error $\mathcal{E}^\mathrm{stable}$ compared to the combined map cycle error $\mathcal{E}^{\F_{\text{MF}} \circ \mathcal{I}_{\text{D}}}$ (bottom row). (Left) In $L_\infty$, worst-case HF radius and velocity predictions $\F_\text{MF}(\hat{\mathbf{T}}_r)$ from estimated drives against $\F_\text{MF}(\mathbf{T}_r)$ (top) and true HF radius and velocity (bottom). Test-set error distributions for radius \revtwo{in $[mm]$} (red) and velocity \revtwo{in $[mm/ns]$} (blue), with denoted $5^\text{th}$, $50^\text{th}$, and $95^\text{th}$ percentiles, for $\mathcal{E}^\mathrm{stable}$ and $\mathcal{E}^{\F_{\text{MF}} \circ \mathcal{I}_{\text{D}}}$ are provided via $L_\infty$ (middle) and relative \revtwo{(nondimensional)} $L_1$ (right) metrics.}
\label{fig:dr_recon_err1}
\end{figure}

Given drive estimates $\hat{\mathbf{T}}_r$ (from $\mathcal{I}_\text{D}$) for $\mathbf{T}_r$ over the test set $\D_{\text{HF}}^{\text{Inv, test}}$, we investigate the forward model's cycle-consistency by evaluating the MF surrogate at these estimated drives, $\F_\text{MF}(\hat{\mathbf{T}}_r)$. The resulting surrogate forward interface trajectories are compared with (a) the surrogate's predictions from corresponding true drives, $\F_\text{MF}(\mathbf{T}_r)$ (\Cref{fig:dr_recon_err1}, top row), and (b) simulated HF interface trajectories $\x^\text{HF}=(\mathbf{R}_i^\text{HF},\mathbf{V}_i^\text{HF})$ from true drives (\Cref{fig:dr_recon_err1}, bottom row), generated as described in \Cref{sec:data}. For each $n\in\mathcal{D}_\text{HF}^\text{Inv, test}$, we denote these errors by
\begin{align}\label{eq:consist}
    \mathcal{E}_n^\mathrm{stable}:= \left|\left|\F_\text{MF}(\hat{\mathbf{T}}_{r,n})- \F_\text{MF}(\mathbf{T}_{r,n})\right|\right|,
    \quad \text{and} \quad
     \mathcal{E}_n^{\F_{\text{MF}} \circ \mathcal{I}_{\text{D}}}:= \left|\left|\F_\text{MF}(\hat{\mathbf{T}}_{r,n})- \mathbf{x}_n^\text{HF}\right|\right|
\end{align}
respectively. Both of these errors are cycle-consistency diagnostics, where $\mathcal{E}^{\F_{\text{MF}} \circ \mathcal{I}_{\text{D}}}$ tells us how well the combined mapping $\F_\text{MF}\circ\mathcal{I}_\text{D}$ infers the true interface trajectories $\x^\text{HF}$, while $\mathcal{E}^\mathrm{stable}$ tells us how sensitive the forward model $\F_\text{MF}$ is to error introduced in the inverse model $\mathcal{I}_\text{D}$. In \Cref{fig:dr_recon_err1} we see that $\mathcal{E}^\mathrm{stable}$ is smaller than $\mathcal{E}^{\F_{\text{MF}} \circ \mathcal{I}_{\text{D}}}$ in distribution; however, they are both are on the same scale as $\F_\text{MF}$'s error in \Cref{fig:HF_error}. This indicates that most of the cycle error is coming from the forward model's own imperfection/generalization error, and that $\F_\text{MF}$ is fairly insensitive to the inverse model's error. This likely arises both due to (a) insensitivity/degeneracy of the true inverse problem, where different temperature drives can produce similar interface dynamics, and (b) the smoothing of sharp transitions by RNNs \cite[Ch.~10]{goodfellow2016deep} combined with the causal nature of $\F_\text{MF}(\hat{\mathbf{T}}_r)$, which attenuates late-time drive perturbations from affecting early- and mid-time predictions. This can be seen by noting the worst case and distributional error in cycle consistency in \Cref{fig:dr_recon_err1} is noticeably smaller than that for just the inverse problem of temperature drive predictions in \Cref{fig:dr_error1}.

\subsection{Optimizing drive from optimal snapshots -- sparse-time networks}
\label{subsec:sparse-nn}

We consider two inverse models $\mathcal{I}_\text{S}^R$ and $\mathcal{I}_\text{S}^{R,V}$ that map just radius $\mathbf{R}_i^\text{HF}$ and both radius and velocity $(\mathbf{R}_i^\text{HF},\mathbf{V}_i^\text{HF})$, respectively, observed at a sparse set of $N_s=4$ times to the principal components of the temperature drive. Our goal is to learn not only a point‐estimate of $\hat{\mathbf{T}}^\text{PCA}_r$ for $\mathbf{T}^\text{PCA}_r$, but also an interpretable selection of the $N_s=4$ most informative (sample-dependent) time steps in the dynamics that influence the estimation. For both models, we propose a single‐stage, end‐to‐end differentiable architecture (inspired by soft selection in \cite{kalchbrenner2014,lin2017structured}), which consists of (a) a linear score layer that softly selects $N_s$ time steps via a temperature‐controlled \texttt{SoftMax}, and (b) an MLP that maps the resulting weighted summaries to the $N_d=4$ principal component output. This architecture combines the interpretability of hard time step selection with the flexibility of a differentiable end‐to‐end model, allowing us to recover both the principal components and the most informative temporal locations in a single pass. Details are provided in \Cref{app:inv2}. Both networks are trained via the MSE loss with same optimizer, learning rate scheduler, and early stopping criterion that was used for the LF and HF modules of the surrogate and the dense-time inverse problem. 

\begin{figure}[h!]
	\centering  
  	\includegraphics[width=.29\linewidth]{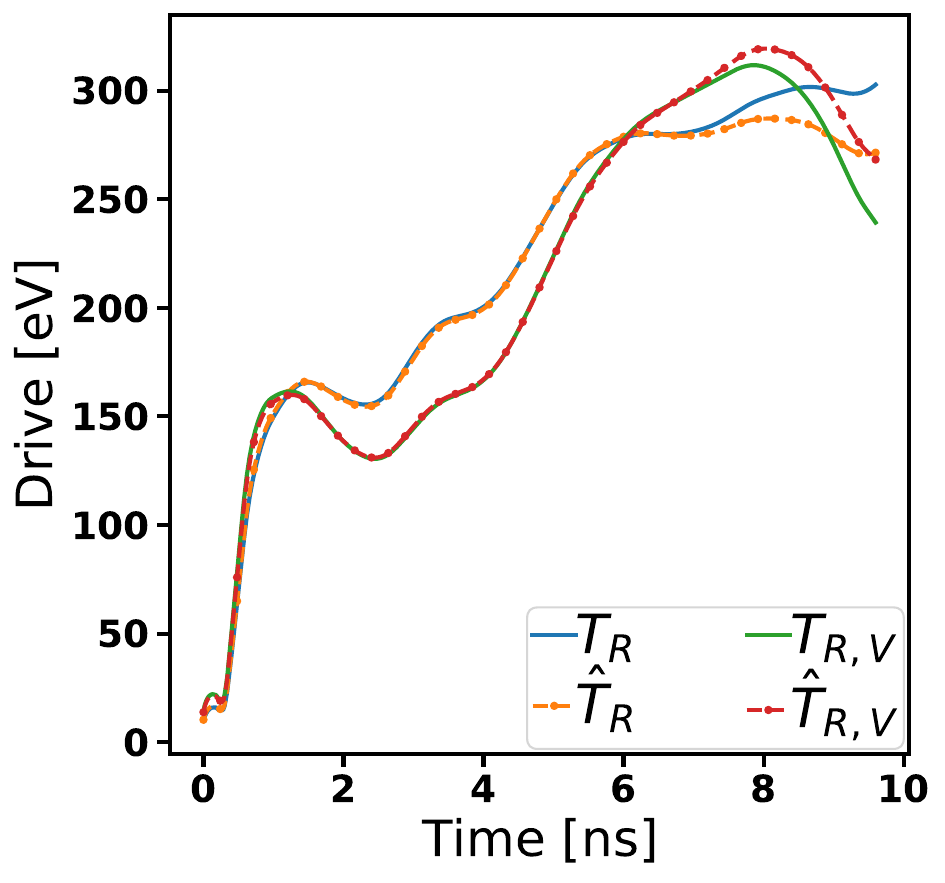}
  	\includegraphics[width=.33\linewidth]{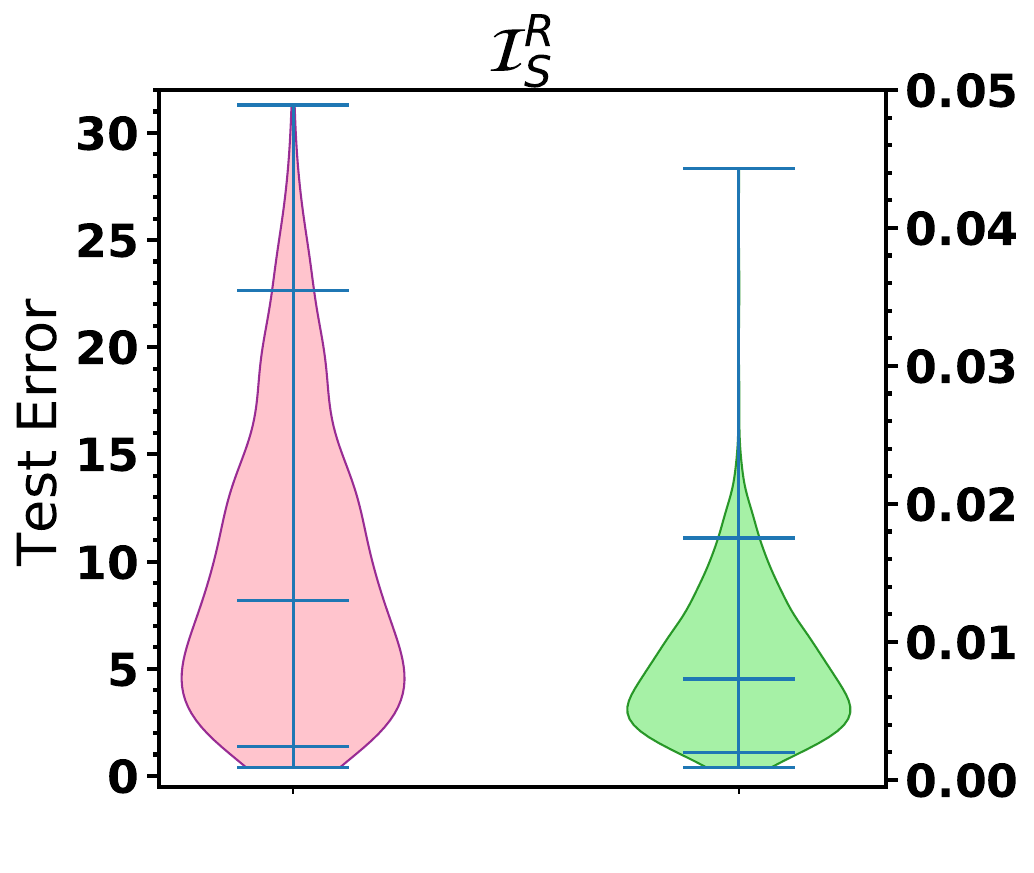}
  	\includegraphics[width=.33\linewidth]{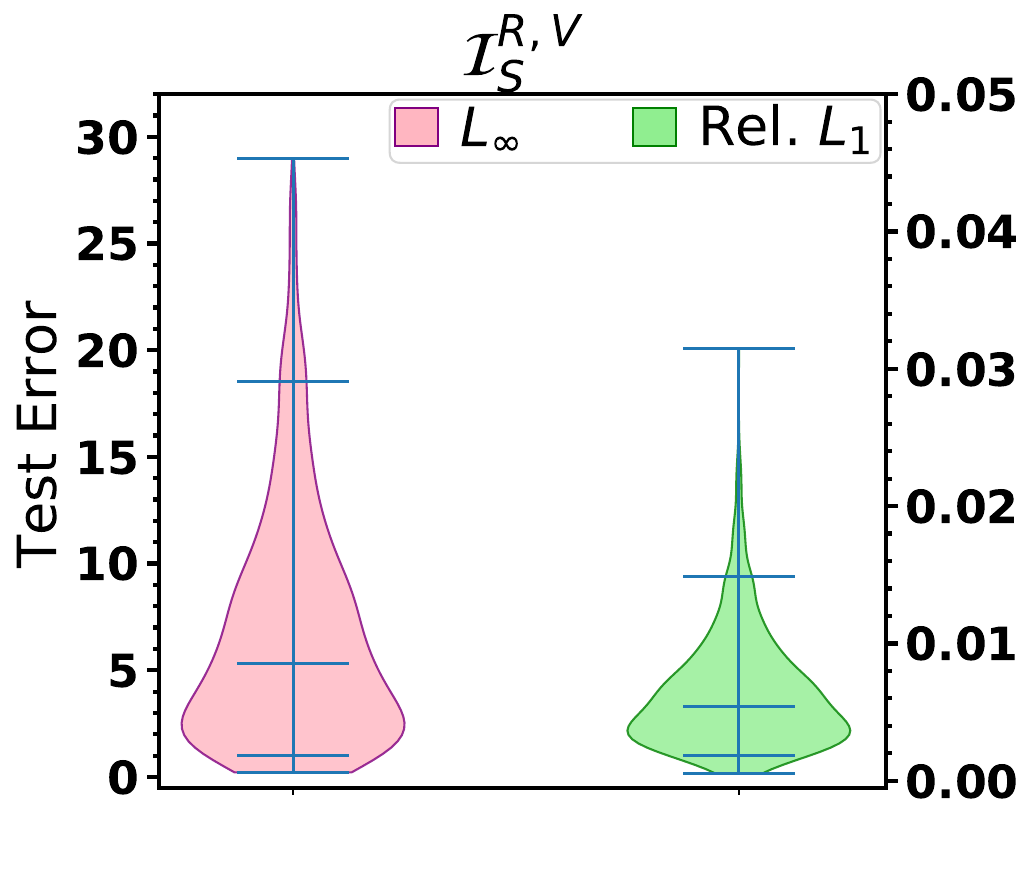}
\caption{(Left) Worst-case drive test-set predictions (in $L_\infty$) for inverse models $\mathcal{I}_\text{S}^{R}$ and $\mathcal{I}_\text{S}^{R,V}$. Test error distributions via $L_\infty$ \revtwo{in $[eV]$} (pink) and relative $L_1$ \revtwo{(nondimensional)} (green) metrics (with denoted $5^\text{th}$, $50^\text{th}$, and $95^\text{th}$ percentiles) are provided for $\mathcal{I}_\text{S}^{R}$ (middle) and $\mathcal{I}_\text{S}^{R,V}$ (right).}
\label{fig:dr_error2}
\end{figure}

\Cref{fig:dr_error2} displays the worst-case test-set predictions (in $L_\infty$) for the drive for both networks $\mathcal{I}_\text{S}^R$ and $\mathcal{I}_\text{S}^{R,V}$, which both occur at $t = t_f = 9.6\,[ns]$, i.e., the final time. The corresponding worst-case $L_\infty$ errors for $\mathcal{I}_\text{S}^R$ and $\mathcal{I}_\text{S}^{R,V}$ are approximately $31.3\,[eV]$ and $28.9\,[eV]$, while the median-case errors are approximately $8.2\,[eV]$ and $5.3\,[eV]$, respectively. The full $L_\infty$ test-error distribution is provided (pink) as well as its counterpart in the relative $L_1$ metric (green). The median and maximum relative $L_1$ errors are $0.7\%$ and $4.4\%$ for $\mathcal{I}_\text{S}^R$ and $0.5\%$ and $3.1\%$ for $\mathcal{I}_\text{S}^{R,V}$, respectively. Additionally, the variance-weighted $R^2$ coefficients over the test set are $0.96$ and $0.972$ for $\mathcal{I}_\text{S}^R$ and $\mathcal{I}_\text{S}^{R,V}$, respectively. Hence, including velocity in addition to radius as input to the network does make a statistically significant improvement to drive estimation, albeit a small one, for a fixed number of $N_s=4$ heads. Arguably, the more remarkable feat is how well inference using four time snapshots with $\mathcal{I}_\text{S}^{R,V}$ compares to using the full time history with $\mathcal{I}_\text{D}$ (see \Cref{fig:dr_error1}), where the test error median approximately differs by only $0.8\,[eV]$ and $0.1\%$ in $L_\infty$ and relative $L_1$, respectively. There is more deviation at the tails of the distributions, but, overall, they are fairly similar. The cycle-consistency errors defined in \cref{eq:consist} corresponding to estimated drives $\hat{\mathbf{T}}_r$ from both $\mathcal{I}_\text{S}^{R}$ and $\mathcal{I}_\text{S}^{R,V}$ are similar (albeit slightly larger) to those seen in \Cref{fig:dr_recon_err1} for $\mathcal{I}_\text{D}$, but are not included for brevity. 

\begin{figure}[h!]
	\centering  
  	\includegraphics[width=.35\linewidth]{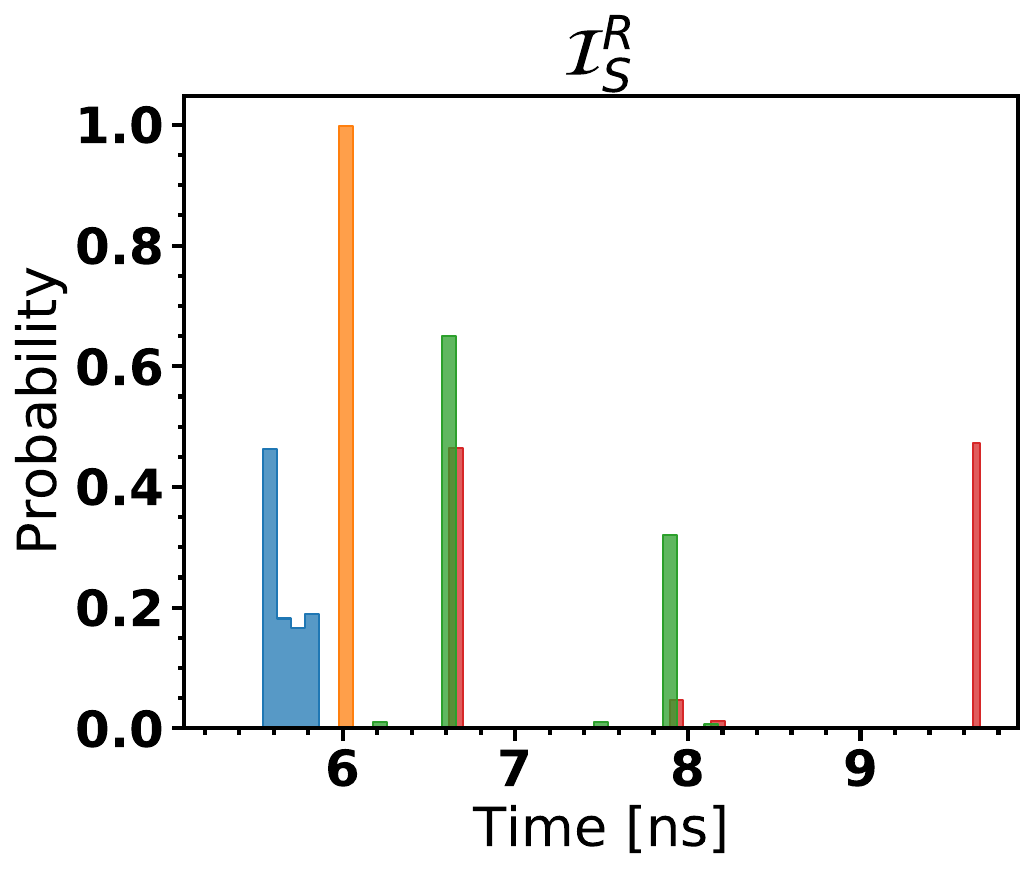}
    \includegraphics[width=.35\linewidth]{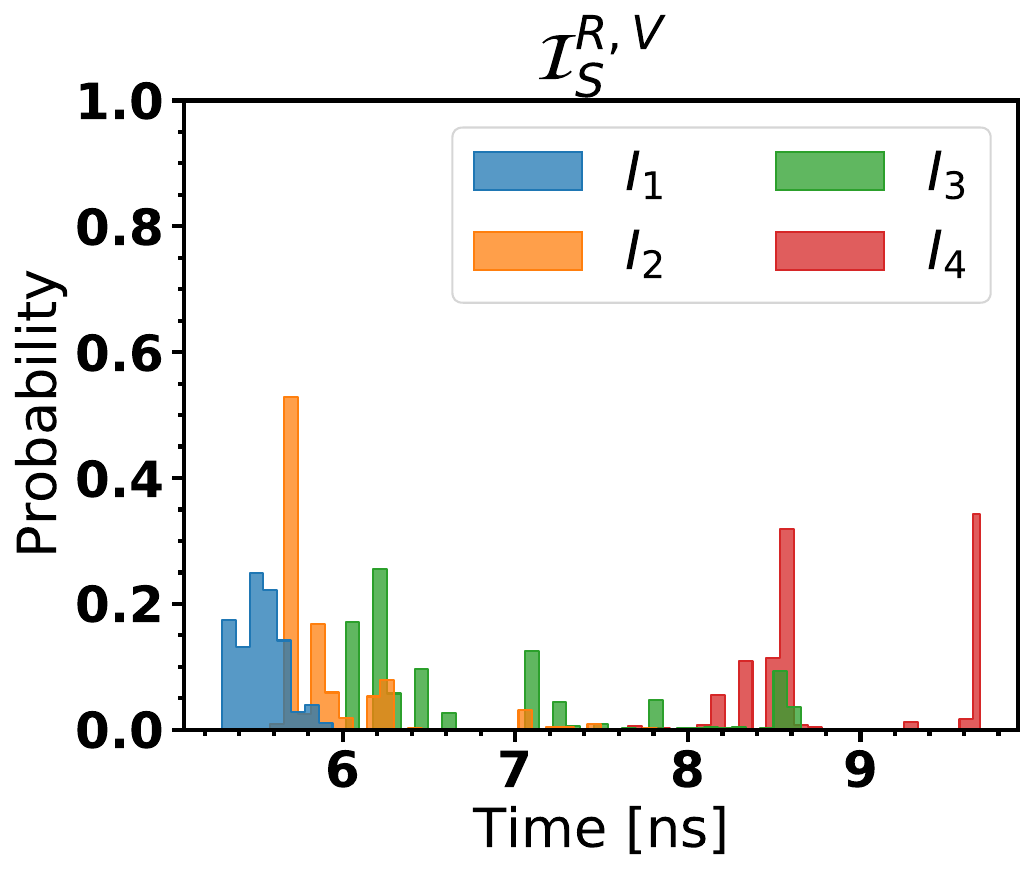}
    \\
    \includegraphics[width=.37\linewidth]{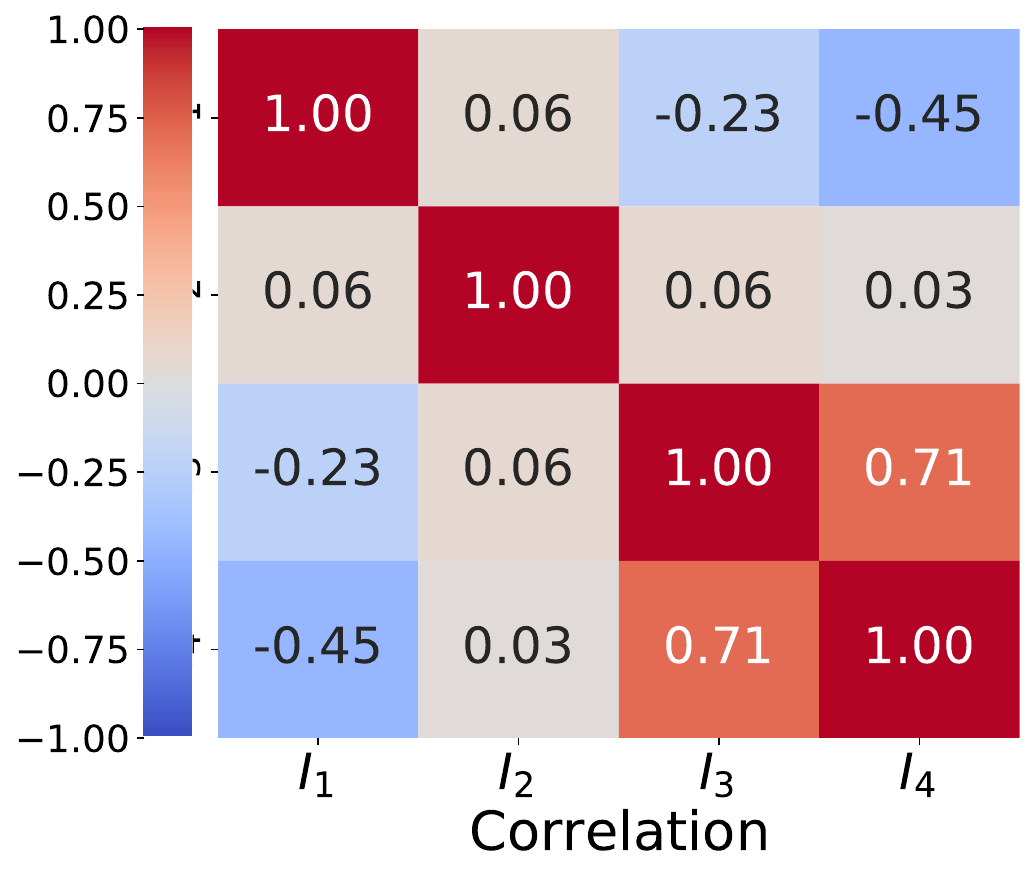}
    \includegraphics[width=.37\linewidth]{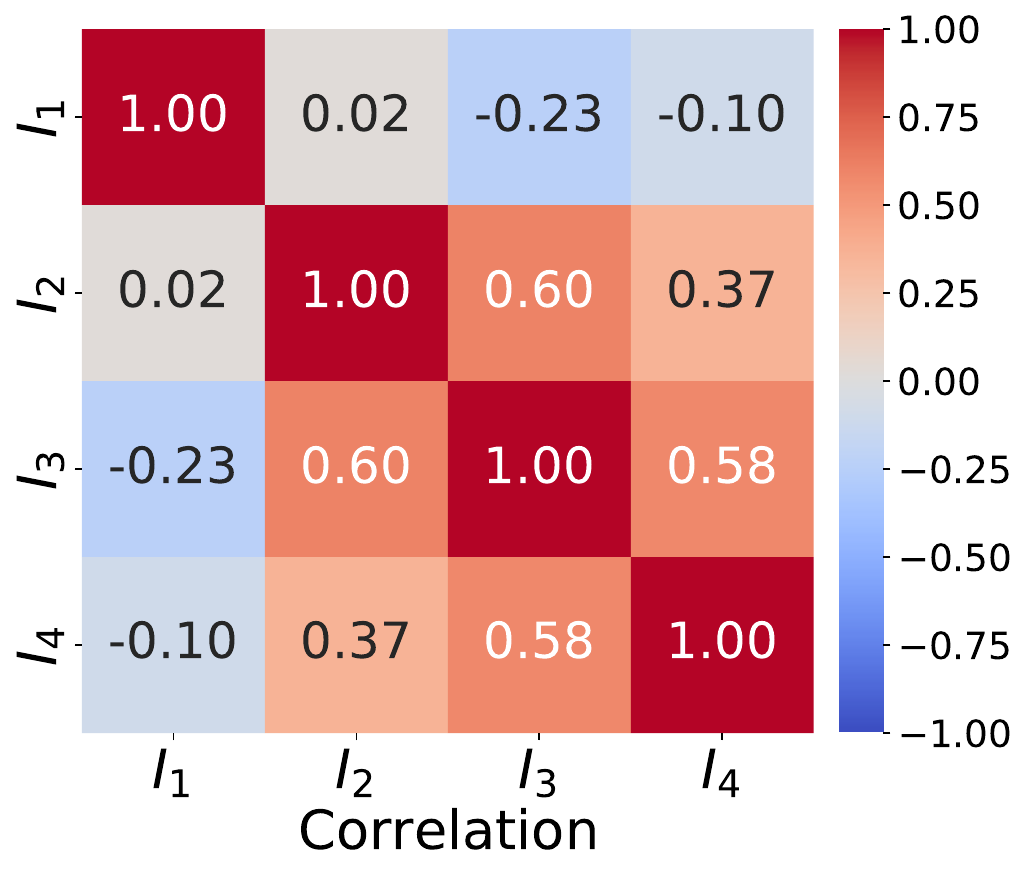}
\caption{Test set histograms and Pearson correlation coefficients of learned times corresponding to the indices $\mathbf{I}$ for networks $\mathcal{I}_\text{S}^{R}$ (left) and $\mathcal{I}_\text{S}^{R,V}$ (right). Note, a handful of bins ranging from $t\approx4\,\text{to}\,5\,[ns]$ have been omitted from the histograms since they contain only a few values and are not visible.}
\label{fig:dr_times}
\end{figure}

\Cref{fig:dr_times} displays the distribution of the $N_s=4$ learned times over the test set for both sparse-time networks. We immediately see for $\mathcal{I}_\text{S}^{R}$, the first two heads of the network choose early times (blue and orange), which typically correspond to the initial acceleration and initial \revthree{ballistic} (i.e., velocity plateau) phases of the implosion, respectively. The latter corresponds to early time when the radius in \cref{eq:bk-control} has nearly linear dynamics. The third head (green) largely alternates between a middle time and a mid-to-late time, while the fourth head (red) alternates between the same middle time and a very late time. \revone{Relating these times precisely to physics is complicated by the fact that these ``optimal times" are ensemble averaged and therefore precise attribution to a precise physical phenomena is not possible. However, the times indicated correspond to approximately shock breakout at the inner surface, onset of significant acceleration, peak velocity, and bang-time.} The network has not been constrained to choose unique times, and heads three and four both pick the middle time $t\approx6.7\,[ns]$ for nearly 40\% of the distribution, which also results in a moderately high Pearson correlation coefficient. Hence, there is redundant information being passed to the MLP for several of the test cases, indicating some radii can be explained sufficiently well by only a few times. We note that $t\approx6.7\,[ns]$ is the only time in $\mathbf{t}$ where the network heads learn a non-unique time. On the other hand, when velocity is fed into the network $\mathcal{I}_\text{S}^{R,V}$, the distribution of learned times is much more diverse even though there exists moderate correlation. In fact, the heads of $\mathcal{I}_\text{S}^{R,V}$ choose unique times for all test cases. The two closest times are always the two earliest times, with the distance between the two ranging from $0.1$ to $0.4\,[ns]$. We also note the average distances between times two and three and times three and four are $0.9$ and $2\,[ns]$, respectively, highlighting how the optimal data sampling is non-uniform in time. Lastly, both networks learn a \texttt{SoftMax} temperature $\hat\gamma<1$, indicating sharp feature selection. In particular, $\hat\gamma\approx0.64$ for $\mathcal{I}_\text{S}^{R}$ and $\hat\gamma\approx0.22$ for $\mathcal{I}_\text{S}^{R,V}$. This produces weights $\sigma_{j,t}^{(n)}$ comprised of a single dominant mode in time for each head $j\in\{1,\dots,N_s\}$, ensuring that $\mathbf{I}^{(n)}$ is indeed a good proxy for optimally selected times.

\begin{remark}[Global time selection]
    When using the forward model and sparse-time inverse models for experimental design, it may be more practical in some settings to have the networks select a set of global/fixed optimal times in place of dynamic/sample-dependent selection. In that case, instead of the per-head scores being updated from the input via the linear map \(W_{\mathrm{score}}\), a score matrix can be treated as a parameter of the network. In the network's forward pass, it is used to directly compute the weights/probabilities via the \texttt{SoftMax}, where the input is used only for computing the features in \cref{eq:soft_times}. Hence, the score matrix is only updated via backpropagation. For example, using the exact same MLP networks described above for $\mathcal{I}_\text{S}^{R}$ and $\mathcal{I}_\text{S}^{R,V}$, where \(W_{\mathrm{score}}\) is replaced with a score matrix network parameter, results in global times $\{5.3, 6.2, 8.2, 8.8\}$ and $\{4.3,5.3,7.8,8.8\}\,[ns]$, respectively. However, there is naturally some loss in accuracy from this more rigid framework. As a brief comparison, the variance-weighted $R^2$ reduces from $0.96$ to $0.95$ for $\mathcal{I}_\text{S}^{R}$ and from $0.972$ to $0.958$ for $\mathcal{I}_\text{S}^{R,V}$.
\end{remark}

\section{Limitations}
\label{sec:limitations}

\revone{While the results are promising, our study has several limitations:}
\revone{
\paragraph{Restriction to 1D capsule models.}
All training and evaluation in this work was performed on 1D capsule simulations. The proposed multi-fielity design transfers information across radiation group resolution \emph{within the same physics model}, but scalability to 2D/3D geometries—with multi-mode radiation-hydrodynamic instabilities, has not yet been validated. Extending to higher dimensions will require new HF datasets, appropriate LF $\to$ HF mappings, and careful handling of trajectory definitions (e.g., interface extraction on evolving meshes).}
\revone{
\paragraph{Dependence on surrogate accuracy in under-sampled tails.}
Despite strong median errors, accuracy degrades in under-sampled tail regimes (e.g., very high peak $T_r$ and late-time sharp deceleration), where the MF residual learner depends on LF predictions. This introduces sensitivity to data scarcity and covariate shift. Mitigations (left to future work) include active sampling for tail coverage, tail-aware loss reweighting, and error-aware controllers in the embedded ODE.}
\revone{\paragraph{Distribution shift across drive families.}
We train on perturbations of a single baseline; zero-shot generalization to substantially different baselines (timings/shock counts) will require either expanded training coverage or modest fine-tuning with LF simulations plus limited HF exemplars.}

\section{Conclusions}
\label{sec:conclusions}
We have presented a comprehensive framework for multi-fidelity surrogate modeling and inverse reconstruction of inertial confinement fusion (ICF) capsule dynamics, leveraging architectures with embedding physics, causal sequence learning, and reduced-order modeling. The proposed surrogate system provides a unified approach to forward and inverse modeling of the deuterium-tritium (DT) shell interface under varying radiation drive profiles, with predictive capabilities that span low- and high-fidelity simulation regimes.
Our approach is built on an embedded, parameterized ODE model of shell dynamics, which enforces physical structure and causality throughout the learning pipeline. The low-fidelity (LF) surrogate maps temporally-resolved radiation drive profiles $T_r(t)$ to control signals $P(t)$ that parameterize the embedded ODE. This mapping is implemented via a causal neural architecture comprising convolutional, LSTM, MLP, and linear skip layers. The LF model is trained on 4000 simulations with a reduced (3-group) radiation model. A high-fidelity (HF) residual network augments this LF controller using a limited training set of 300 67-group radiation-hydrodynamics simulations, learning a data-driven correction to the LF prediction. This two-tier surrogate maintains efficiency while dramatically improving accuracy.

We also proposed a family of inverse models that recover the drive $T_r(t)$ from observable interface dynamics $\x(t) = (R_i(t), V_i(t))$. The dense-time inverse model processes full implosion trajectories and maps them to a compact, 4-dimensional PCA representation of the drive. In contrast, the sparse-time inverse model jointly learns optimal sampling times and a mapping from a small number of snapshot observations (at most four) to the same PCA representation. The sparse-time model adapts to different combinations of inputs (e.g., radius-only vs. radius + velocity), enabling flexible deployment across diagnostic-limited settings.

All components of the surrogate framework were evaluated using a combination of worst-case trajectory reconstructions, distributional error metrics ($L_\infty$ and relative $L_1$), and a cycle-consistency check. The latter validates that reconstructed drives from the inverse models, when passed through the forward surrogate, reproduce the original shell dynamics with minimal deviation. This provides strong empirical evidence of physical coherence and architectural fidelity.
Together, these results establish a scalable and interpretable path forward for data-driven plasma science, demonstrating how operator learning, causal architectures, and physical inductive bias can be effectively integrated to accelerate discovery, design, and diagnostics in high-energy-density systems.

\revone{Future work will extend the framework developed here to more complex regimes. For higher dimensional problems, we will focus on extending surrogates to capture multimode perturbations, low-mode asymmetries, and burn physics (e.g., alpha heating feedback), and also validate scalability to 2D/3D coupled rad-hydro problems. Because our surrogate already provides $R_i(t)$ and $V_i(t)$ through peak velocity and into early deceleration, a natural next step is to couple the shell surrogate to a dynamic hotspot model. We will also focus on closed-loop optimization pipelines that use the causal surrogate for on-the-fly experimental design (e.g., pulse shaping, snapshot scheduling) with compute-budget awareness. \revthree{Additional extensions will also examine the use of higher-fidelity radiation transport models, e.g., Sn transport or IMCC. There is also the possibility of using special keyhole targets and liquid D$_2$ (or thick DT ice) as transparent fuel emulators to enable the tracking of the shock-front velocity inside the fuel as it transits the capsule.} Finally, we plan to integrate the causal surrogte with UQ frameworks (e.g., conformal and risk-aware objectives) to deliver calibrated coverage, reliability under distribution shift, and robust optimization over model and data uncertainty.}
\bibliographystyle{plainnat}   %
\bibliography{references}      %

\section*{Acknowledgments}
We would like to thank David Meyerhofer for many useful conversations, reading the manuscript and making many suggestions.

This work was supported by the Laboratory Directed Research and Development program at Los Alamos National Laboratory. LA-UR-25-29020.

\section*{Author contributions}

TEM, BSS, and MLK developed the problem formulation and conceptual framework. JRH and MLK generated high fidelity simulation data. TEM developed and trained all ML models and analyzed data and performance. TEM, BSS, and MLK wrote the manuscript.  

\section*{Competing interests}

The authors declare no competing interests.

\section*{Data availability statement}

The data and code used to generate the results in this study are not currently publicly available, but may be obtained from the authors upon reasonable request.

\clearpage
\pagenumbering{gobble}
\begin{center}\LARGE Supplementary Material for ``Causal Multi-fidelity Surrogate Forward and Inverse Models for ICF Implosions''\end{center}
\appendix

\section{ODE Derivation of Imploding Incompressible Shell}
\label{app:book-derivation}
We derive a first-order ODE system for the inner and outer radii $(R_i,R_o)$ and velocities $(V_i, V_o)$ of an imploding 1D incompressible shell from \cite{Book1987}. Here, the subscript $i$ denotes the shell's inner surface while the subscript $o$ denotes the outer surface. 

The shell is assumed to have uniform density $\rho(R) \equiv \bar{\rho}$. Mass is defined as
\begin{align}\label{eq:bk-mass}
    M := 4\pi\int_{R_i}^{R_o} \rho(R)R^2dR = \frac{4\pi}{3}\bar{\rho}(R_o^3 - R_i^3) \coloneqq \frac{4\pi}{3}\bar\rho R_c^3.
\end{align}
By consequence of conservation of mass, $dM/dt \equiv 0$ and we have
\begin{align}\label{eq:bk-R-relation}
    R_o^2\dot{R}_o - R_i^2\dot{R}_i = 0.
\end{align}
Thus, there is a square-radius dependence between the velocities of the inner and outer surface. Specifically,
\begin{align} \label{eq:bk-V-relation}
    V_i = \left(\frac{R_o}{R_i}\right)^2V_o \qquad \text{or} \qquad V_o = \left(\frac{R_i}{R_o}\right)^2V_i,
\end{align}
where $V_i:=\dot{R}_i$ and $V_o:=\dot{R}_o$.
Energy follows from a standard $E = \tfrac{1}{2}M\dot{R}^2$ argument, but we treat the velocity $\dot{R}$ pointwise and integrate against mass over density and the sphere. The mass integral is a result of integrating $\int \rho(R) d\V$ over shell volume $\V \sim (4\pi/3)R^3$, so $d\V = 4\pi R^2dR$. Specifically, we have total kinetic energy as a function of time
\begin{subequations} \label{eq:bk-w}
\begin{align}
    W(t) := \frac{1}{2}\int \rho(R)\dot{R}^2 d\V 
    &= 2\pi\bar\rho\int_{R_i}^{R_o} R^2\dot{R}^2 dR 
     = 2\pi\bar\rho\int_{R_i}^{R_o} (R^2\dot{R})^2\tfrac{1}{R^2} dR \\
    &= 2\pi\bar\rho(R_i^2\dot{R}_i)^2\int_{R_i}^{R_o} \tfrac{1}{R^2} dR 
    = 2\pi\bar\rho \dot{R}_i^2 R_i^3(1-R_i/R_o),\label{eq:bk-tot-energy}
\end{align}
\end{subequations}
where the penultimate term in \cref{eq:bk-w} follows from \cref{eq:bk-R-relation} such that $R^2\dot{R}$ is constant across the shell for incompressible flow.

With no external forces, conservation of energy corresponds to $\tfrac{d}{dt}W(t) \equiv 0$, which is the basis for the derivation in \cite{Book1987}. Hence, $W$ is constant such that
\begin{align} \label{eq:w0}
    W\equiv W_0 := 2\pi\bar\rho V_i^2(0)R_i^3(0)(1-R_i(0)/R_o(0))
\end{align}
Rearranging \cref{eq:bk-tot-energy} we have
\begin{align}\label{eq:bk-tot-energy2}
    \dot{R}_i^2 = \frac{W}{2\pi\bar\rho R_i^3(1-R_i/R_o)}.
\end{align}
Differentiating both sides and significant algebra \cite[Eq. 6]{Book1987} yields a 1D model for an incompressible imploding shell into vacuum:
\begin{align}\label{eq:bk-dtot-energy1}
    \dot{V}_i =\frac{-W_0}{4\pi\bar\rho R_i^4} \cdot\left[3 + 2\frac{R_i}{R_o} + \left(\frac{R_i}{R_o}\right)^2\right].
\end{align}

To get an equivalent equation in $R_o$, from \cref{eq:bk-V-relation} $V_i^2 = (R_o/R_i)^4V_o^2$. Substituting into \cref{eq:bk-tot-energy2} we have
\begin{align}\label{eq:bk-tot-energy3}
    V_o^2 = \frac{-W_0}{2\pi\bar\rho R_o^3(1-R_o/R_i)}.
\end{align}
Notice that this is equivalent to \cref{eq:bk-tot-energy2} with a sign change and the roles of $R_i$ and $R_o$ swapped.
An analogous derivation as \cref{eq:bk-dtot-energy1} yields
\begin{align}\label{eq:bk-dtot-energy2}
    \dot{V}_o = \frac{W_0}{4\pi\bar\rho R_o^4} \cdot\left[3 + 2\frac{R_o}{R_1} + \left(\frac{R_o}{R_i}\right)^2\right].
\end{align}
Together, this yields a coupled second-order system of ODEs.
In this setting, we can also express $R_o$ directly as a function of $R_i$ by enforcing a fixed mass and incompressible material, thus ensuring that the imploded shell radius is defined by
\begin{align}
    R_c^3 := R_o^3 - R_i^3. 
\end{align}
Solving for $R_o = \sqrt[3]{R_c^3 + R_i^3}$ and plugging into \cref{eq:bk-dtot-energy1} gives a closed form for an ODE system only in $R_i$,
\begin{align}
    \ddot{R}_i =\frac{-W_0}{4\pi\bar\rho R_i^4} \cdot\left[3 + \frac{2R_i}{\sqrt[3]{R_c^3 + R_i^3}} + \left(\frac{R_i}{\sqrt[3]{R_c^3 + R_i^3}}\right)^2\right],
\end{align}
which has first-order form
\begin{align} \label{eq:app-bk}
    \dot R_i &= V_i, \notag \\
    \dot V_i &= \frac{-W_0}{4\pi\bar\rho R_i^4} \cdot\left[3 + \frac{2R_i}{\sqrt[3]{R_c^3 + R_i^3}} + \left(\frac{R_i}{\sqrt[3]{R_c^3 + R_i^3}}\right)^2\right].
\end{align}

In order to manipulate an implosion, we introduce a source term to \cref{eq:app-bk} in the form of a power function $P(t)$ such that the total kinetic energy is given by
\begin{align} \label{eq:bk-wt}
    W \equiv 2\pi\bar\rho V_i^2(t)R_i^3(t)(1-R_i(t)/R_o(t)) = W_0 + \int_0^t P(s)\,ds.
\end{align}
We proceed as before by differentiating both sides of \cref{eq:bk-tot-energy}; however, $W$ is now time-dependent. We have
\begin{align} \label{eq:bk-p-derivation}
    2\dot{R}_i\ddot{R}_i & = \frac{-W}{2\pi\bar\rho (R_i^3(1-R_i/R_o))^2}\frac{\d}{\d t}\left(R_i^3 - R_i^4/R_o\right) + \frac{\dot{W}}{2\pi\bar\rho R_i^3(1-R_i/R_o)}.
\end{align}
The first term in the right-hand side of \cref{eq:bk-p-derivation} is identical to that seen in the derivation of \cref{eq:bk-dtot-energy1}, and the second term reduces to $P\dot{R}_i^2/W$. Dividing both sides by $2\dot R_i$, applying the same algebra used for \cref{eq:bk-dtot-energy1}, and re-writing as a first-order system yields
\begin{align} \label{eq:app-bk-p}
    \dot R_i &= V_i, \notag \\
    \dot V_i &= \frac{-W}{4\pi\bar\rho R_i^4} \cdot\left[3 + \frac{2R_i}{\sqrt[3]{R_c^3 + R_i^3}} + \left(\frac{R_i}{\sqrt[3]{R_c^3 + R_i^3}}\right)^2\right] + \frac{PV_i}{2W}.
\end{align}

\section{Adjoint Method for ODE Constrained Optimal Control}
\label{app:adjoint-control}

We want to solve
\begin{align} \label{eq:control-problem}
    &\min_\u J(\x,\u) := \int_{t_0}^{t_f} f(\x,\u,t) \,dt \\
    s.t. &\begin{cases}
            \bar\h(\x,\dot\x,\u,t)=0 \notag \\
            \g(\x(t_0),\u) = 0. \notag 
         \end{cases}
\end{align}
For a gradient-based optimization routine, we need to compute the total derivative of $F$ with respect to the control $\u$:
\begin{align}
    D_\u J(\x,\u) = \int_{t_0}^{t_f} [\d_\x f D_\u \x + \d_\u f]\,dt.
\end{align}
We derive a first-order ODE system for the adjoint $\lam$ that is instrumental in calculating the gradient. Its benefit is that the total work of computing $F$ and its gradient is approximately equivalent to solving only two ODE systems. We reproduce the derivation from \cite{Cao2003}.

We start by computing the Lagrangian:
\begin{align} \label{eq:lagrangian}
    \mathcal{L} := \int_{t_0}^{t_f} \left[f(\x,\u,t)+\lam^\top\bar\h(\x,\dot\x,\u,t)\right]dt +\muu^\top\g(\x(t_0),\u),
\end{align}
where $\lam(t)$ is the costate/Lagrangian multiplier and $\muu$ is a multiplier for the initial conditions. Since $\bar\h\equiv\g\equiv0$ by construction, we are free to set values of $\lam$, $\muu$, and $D_\u\mathcal{L}=D_\u J$. Taking the total derivative of \cref{eq:lagrangian} gives
\begin{align} \label{eq:lagrang-gradient}
    D_\u\mathcal{L} &= \int_{t_0}^{t_f} \left[ \d_\x f D_\u \x + \d_\u f +\lam^\top \left( \d_\x \bar\h D_\u \x + \d_{\dot\x}\bar\h D_\u \dot\x + \d_\u \bar\h\right)\right]dt \\
    &+ \muu^\top(\d_{\x(t_0)} \g D_\u \x(t_0) + \d_\u \g). \notag
\end{align}
Integrating by parts removes the $D_\u \dot\x$ term:
\begin{align}
    \int_{t_0}^{t_f} \lam^\top \d_{\dot\x}\bar\h D_\u \dot\x\,dt = \lam^\top \d_{\dot\x}\bar\h D_\u \x \Big|_{t_0}^{t_f} - \int_{t_0}^{t_f} \left[\dot\lam^\top \d_{\dot\x}\bar\h + \lam^\top D_t\d_{\dot\x}\bar\h\right] D_\u\x\,dt.
\end{align}
Substituting this into \cref{eq:lagrang-gradient} and grouping terms in $D_\u\x$ and $D_\u\x(t_0)$ yield
\begin{align}
    D_\u\mathcal{L} &= \int_{t_0}^{t_f} \left[\left(\d_\x f + \lam^\top \left(\d_\x\bar\h - D_t\d_{\dot\x}\bar\h\right) -\dot\lam^\top \d_{\dot\x}\bar\h\right)D_\u\x + f_\u +\lam^\top \bar\h_\u\right]dt \\
    &+ \lam^\top\d_{\dot\x}\bar\h \,\,D_\u \x \Big|_{t_f} + \left(-\lam^\top\d_{\dot\x}\bar\h+\muu^\top \g_{\x(t_0)}\right)\Big|_{t_0} D_\u\x(t_0) + \muu^\top \g_\u. \notag 
\end{align}
Since $D_\u\x$ is difficult to calculate, we set $\lam(t_f)=0$ and $\muu^\top = \lam^\top\d_{\dot\x}\bar\h|_{t_0}\g_{\x(t_0)}^{-1}$ to cancel the first two terms outside of the integral. Moreover, we set 
\begin{align} \label{eq:adjoint-eq}
    \d_\x f + \lam^\top(\d_\x\bar\h - D_t\d_{\dot\x}\bar\h) - \dot\lam^\top\d_{\dot\x}\bar\h = 0.
\end{align}
to avoid $D_\u \x$ in the integrand. Therefore, the algorithm for $D_\u J$ is:
\begin{enumerate}
    \item Integrate $\bar\h(\x,\dot\x,\u,t)=0$ for $\x$ over $t\in(t_0,t_f]$ with initial condition $\g(\x(t_0),\u)=0$.
    \item Integrate \cref{eq:adjoint-eq} for $\lam$ over $t\in(t_f,t_0]$ with terminal condition $\lam(t_f)=0$.
    \item Set 
        \begin{align} \label{eq:dF}
            D_\u J = \int_{t_0}^{t_f} \left[f_\u +\lam^\top\d_\u\bar\h\right]dt +\lam^\top\d_{\dot\x}\bar\h|_{t_0}\,\g_{\x(t_0)}^{-1}\g_\u.
        \end{align}
\end{enumerate}

In the special case that the ODE constraint is in explicit form, i.e., $\bar\h := \dot\x - \h(\x,\u,t)$, we can further simplify. The adjoint equation \cref{eq:adjoint-eq} becomes
\begin{align} \label{eq:adjoint-eq-simplified}
    \dot\lam^\top + \lam^\top\d_\x\h - \d_\x f = 0, \qquad \lam(t_f)=0,
\end{align}
and the gradient \cref{eq:dF} becomes
\begin{align} \label{eq:dF-simplified}
    D_\u J = \int_{t_0}^{t_f} \left[f_\u -\lam^\top\d_\u\h\right]dt +\lam^\top(t_0)\,\g_{\x(t_0)}^{-1}\g_\u.
\end{align}
Moreover, if the initial condition for the ODE constraint does not depend on the control $\u$, the term outside of the integral in \cref{eq:dF-simplified} vanishes.

The gradient for our specific problem reduces to 
\begin{align} \label{eq:bk-dJ}
    D_{p_k} J = -\int_{\tau_k}^{\tau_{k+1}} \lam^\top(t)\,\d_{p_k}\h(t)\,dt
    = -\int_{\tau_k}^{\tau_{k+1}} \lambda_2(t)\,\d_{p_k}h_2(t)\,dt,
\end{align}
where
\[ 
\d_{p_k}h_2(t) = \frac{-(t-\tau_k)}{4\pi\bar\rho R_i^4} \cdot\left[3 + \frac{2R_i}{R_o} + \left(\frac{R_i}{R_o}\right)^2\right] + \frac{\left[\tilde W_k - p_k(t-\tau_k)\right]V_i}{2\tilde W_k^2}, \quad t\in (\tau_k,\tau_{k+1}].
\]
Here, $\tilde W_k \equiv W$ for $t\in (\tau_k,\tau_{k+1}]$, assuming the previous controller coefficients $\{\tilde p_1,\dots,\tilde p_{k-1}\}$ have been computed. That is, $\tilde W_k(t):= W_0 + p_k(t-\tau_k) + \sum_{j=1}^{k-1} \tilde p_j(\tau_{j+1}-\tau_j)$ for $t\in (\tau_k,\tau_{k+1}]$. The costate $\lam$ is the solution to the adjoint system \cref{eq:adjoint-eq-simplified}, where the Jacobian $\d_\x\h$ of the velocity field in \cref{eq:bk-control} with respect to system states $\x$ is needed. The Jacobian is given by 
\begin{align} \label{eq:bk-jacobian}
    \d_\x\h =
    \begin{pmatrix}
        0 & 1 \\
        \frac{\tilde W_k}{2\pi\bar\rho R_i^5}\left[ 6 + \frac{3R_i}{R_o}+ \left(\frac{R_i}{R_o}\right)^2+\left(\frac{R_i}{R_o}\right)^4+\left(\frac{R_i}{R_o}\right)^5\right] & \frac{p_k}{2\tilde W_k}
    \end{pmatrix},
\end{align}
and therefore the adjoint equation \cref{eq:adjoint-eq-simplified} reduces to 
\begin{align*}
    \dot\lambda_1 &= \frac{-\lambda_2\tilde W_k}{2\pi\bar\rho R_i^5}\left[ 6 + \frac{3R_i}{R_o}+ \left(\frac{R_i}{R_o}\right)^2+\left(\frac{R_i}{R_o}\right)^4+\left(\frac{R_i}{R_o}\right)^5\right], \qquad t\in (\tau_{k+1},\tau_k] \notag \\
    \dot\lambda_2 &= -\lambda_1 - \frac{\lambda_2 p_k}{2\tilde W_k} + V_i - V_i^\text{ref},
\end{align*}
with terminal condition $\lam(\tau_{k+1}) = 0$.

\begin{remark} \label{rmk:vel_thresh}
    \revtwo{As described in \Cref{rmk:plots}, there is a delay between the drive injecting energy into the shell and when the DT interface actually begins to move inwardly, resulting in the radius being constant and the velocity zero for the first few nanoseconds of the dynamics.} This means the energy constant is $W_0=0$ in our ODE model, which naturally causes numerical issues at early times due to \cref{eq:bk-control} not being well-defined. To circumvent this, (recalling $V_i\le0$) we fix a threshold $\epsilon>0$, and find the largest knot $\tau_{k^{*}}$ such that $V_i^\text{ref}(t) < -\epsilon$ for all $t\ge \tau_{k^{*}+1}$. In other words, $\tau_{k^*}$ is the knot immediately to left of the time when $V_i^\text{ref}\approx -\epsilon$. We then fix the controller coefficients $\tilde p_k = 0$ for all $k<k^*$ and solve the sequence of control problems \cref{eq:bk-wf-control} starting with $p_{k^*}$ over $(\tau_{k^*},\tau_{k^{*}+1}]$ using the velocity initial condition $V_i(\tau_{k^*}) = -\epsilon$ in the ODE constraint. We fix $\epsilon=10^{-3}$, which gives a fixed $W_0 \approx 9.56\times 10^{-8}$ for the given NIF capsule configuration. This eliminates numerical issues in the control formulation arising from when $V_i$ is near or exactly zero and there is division of small $P$ by small $W$. This does mean that, for any given simulation/reference trajectory $\x_n$ in the data sets $\D_\text{LF}^\text{For}$ or $\D_\text{HF}^\text{For}$, the controlled solution $\tilde\x_n$ at $t=\tau_{k_n^*}$ has error on the order of $\mathcal O(\epsilon)$. However, this corresponds to a velocity error on the order of $1\,[\mu m/ns]$ for our choice of $\epsilon$, which is sufficiently small in practice. \revtwo{Note, given that the LF and HF forward networks are trained and validated in the controller space, this procedure of using the velocity threshold to first left-pad and then compute controllers is done for all targets in both the training and validation splits of the LF and HF data sets. However, at inference time, while the same $W_0>0$ is used in solving \cref{eq:bk-control} for both LF and HF controller predictions, such predictions are not left-padded, as that would require the ``unseen" ground truth trajectories. Instead, \cref{eq:bk-control} is integrated over the entire domain $(0,t_f]$ for all predictions.}
\end{remark}

To solve the coupled adjoint system involves successively solving the forward system \cref{eq:bk-control} over $t\in (\tau_k, \tau_{k+1}]$ and backward/adjoint system over $t\in (\tau_{k+1}, \tau_{k}]$ for costates $\lam(t)$ to update the gradient $D_{p_k}J$ during each optimization iteration of \cref{eq:bk-wf-control}. The adjoint system is discretized on the same dense uniform temporal mesh $\mathbf t$ where our reference trajectories $\x^\text{ref}$ are available. The integrals in \cref{eq:bk-wf-control} are computed via midpoint/trapezoidal numerical integration, and the forward and backward equations are solved with a fully implicit, variable-order backward differentiation scheme, specifically the \texttt{BDF} method in the \texttt{scipy.integrate.solve\_ivp} routine from \texttt{SciPy} \cite{scipy}. Each of the sequential nonlinear programs \cref{eq:bk-wf-control} together with their gradients \cref{eq:bk-dJ} are solved via the \texttt{LBFGS} optimizer in SciPy's \texttt{scipy.optimize.minimize} using default arguments, where the initial guess for $p_k$ is taken to be zero for all $k\in\{1,\dots,N_k-1\}$. In order to keep the space $\mathcal P$ of controllers fixed, $N_k$ is fixed for all simulations for both low- and high-fidelity data. Note, to simplify the numerics, the uniform time step $\Delta t$ corresponding to $\mathbf t$ and the number of uniform knots $N_k$ are chosen so that $\Delta t$ evenly divides $N_k-1$. In particular, four uniform time steps of length $\Delta t$ can be taken in each uniform knot interval $(\tau_k,\tau_{k+1}]$.

\section{Low-fidelity architecture details}\label{app:LF}

\begin{figure}[h!]
	\centering  
    \begin{tikzpicture}[
        node distance=1cm and 1cm,
        module/.style={draw, rounded corners, minimum width=2cm, minimum height=.75cm, thick, fill=#1!20},
        arrow/.style={-{Latex[length=3mm]}, thick},
        skip/.style={draw=orange, thick, dashed},
        font=\small
    ]

    \node[draw,rectangle, fill=white!80!black, inner sep=5pt, minimum width=10pt, minimum height=80pt]
    (input) at (0,0) {$\mathbf{T}_r$};
    \node at ($(input.south)+(0,-0.3)$) {\scriptsize $(N_B, N_t, 1)$};
    
    \node[module=blue] (conv) [right=of input] {Conv1D};
    \node at ($(conv.south)+(0,-0.3)$) {\scriptsize $(N_B, N_k-1, N_\ell)$};
    
    \node[module=purple] (lnorm) [right=of conv] {Norm};
    \node at ($(lnorm.south)+(0,-0.3)$) {\scriptsize $(N_B, N_k-1, N_\ell)$};
    
    \node[module=orange!70!black] (skip1) [above=of lnorm] {Skip$_\text{LSTM}$};
    \node at ($(skip1.south)+(-0.25,-0.5)$) {\scriptsize $\boldsymbol\alpha_1$};
    \node at ($(skip1.north)+(0,+0.3)$) {\scriptsize $(N_B, N_k-1, N_\ell)$};
    
    \node[module=green!50!black] (lstm) [right=of lnorm] {LSTM};
    
    \node[draw,circle,fill=white,inner sep=1pt] (sum1) [right=of skip1, above=1.2cm of lstm] {$+$};
    
    \node[module=yellow!70!black] (mlp) [right=1.2cm of sum1] {MLP};
    \node at ($(mlp.north)+(0,+0.3)$) {\scriptsize $(N_B, N_k-1, 1)$};
    
    \node[module=orange!70!black] (skip2) [below=of lstm] {Skip$_\text{MLP}$};
    \node at ($(skip2.north)+(-0.25,+0.5)$) {\scriptsize $\boldsymbol\alpha_2$};
    \node at ($(skip2.south)+(0,-0.3)$) {\scriptsize $(N_B, N_k-1, 1)$};
    
    \node[draw,circle,fill=white,inner sep=1pt] (sum2) [below=of mlp, right=1.2cm of lstm] {$+$};
    
    \node[draw,rectangle, fill=white!80!black, inner sep=2pt, minimum width=5pt, minimum height=40pt] (output) [right=of sum2] {$\hat{\mathbf{p}}^\text{LF}$};
    \node at ($(output.south)+(0,-0.3)$) {\scriptsize $(N_B, N_k-1)$};

    \draw[arrow] (input) -- (conv);
    \draw[arrow] (conv) -- (lnorm);
    \draw[arrow] (lnorm) -- (lstm);
    \draw[arrow] (lstm) -- (sum1);
    \draw[arrow] (sum1) -- (mlp);
    \draw[arrow] (mlp) -- (sum2);
    \draw[arrow] (sum2) -- (output);

    \draw[skip] (lnorm) |- (skip1.south);
    \draw[skip] (skip1) -| (sum1.west);
    
    \draw[skip] (lstm) |- (skip2.north);
    \draw[skip] (skip2) -| (sum2);

    \draw[thick] ($(input)+(-1.0,3.1)$) rectangle ($(output)+(1.0,-3.1)$);
    
    \end{tikzpicture}
\caption{Architectural diagram for the LF surrogate $\F_\text{LF}$. We have provided output dimensions for each module apart from the LSTM, which is identical to the Conv1D and LayerNorm output dimensions. Here, $N_B$ is the batch size, $N_\ell=192$ is the latent space/hidden unit dimension for the convolution and LSTM, and $N_k=121$ is the number of controller knots.}
\label{fig:LFsurrogate}
\end{figure}
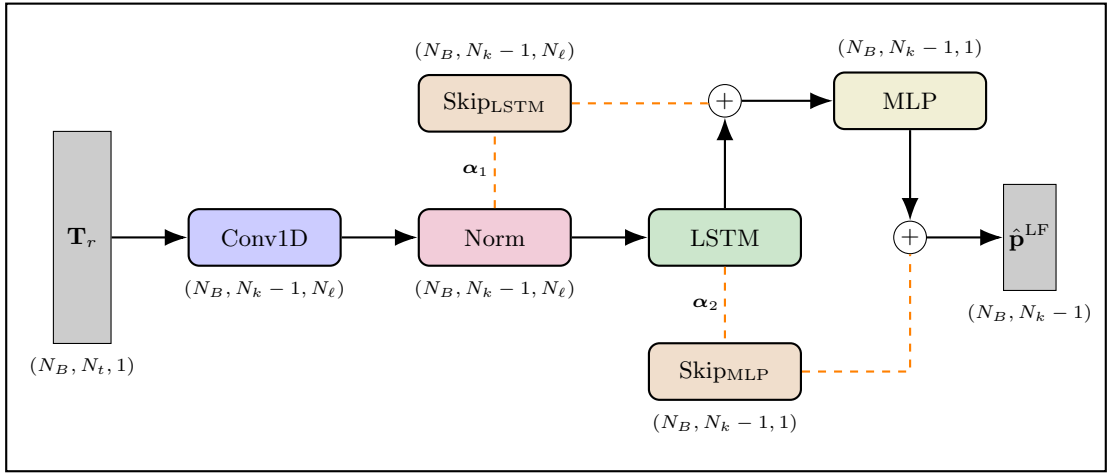

Rather than using fixed downsampling (e.g., strided slicing), we learn a single-layer TCN: a causal 1D convolution $\text{Conv1D}(\mathbf T_r) := W_C * \mathbf T_r \in \mathbb{R}^{(N_k-1)\times N_\ell}$, where
\begin{align*}
\text{Conv1D}[k,j] &= \sum_{m=1}^\mathrm{kernel}W_C[j,m]\cdot T_r[s\cdot k+1-(m-1)]\in\mathbb{R},\quad k\in\{1,\dots,N_k-1\},\,\, j\in\{1,\dots,N_\ell\}, \\
\text{with}\quad T_r[\eta] &:= 0 \quad\text{if}\quad \eta<1 \quad \text{for causality}.
\end{align*}
We set $s:=\text{stride}=4$ to account for $4\cdot \Delta t$ being the length of a knot interval and $\mathrm{kernel}=2s=8$ to avoid over-smoothing while still accounting for the stride. Note, $W_C\in\mathbb{R}^{N_\ell\times\mathrm{kernel}}$ is typically a 3D tensor \cite[Ch.~9]{goodfellow2016deep}, where $N_\ell=192$ is the number of output channels, but we have collapsed the middle dimension since there is only one input channel. The result is one latent feature vector for each knot interval, resulting in the same temporal resolution as the controller coefficients, which is necessary for the LSTM to be aligned in time. A \texttt{LayerNorm} is applied to stabilize feature magnitude before the recurrent stage \cite{Ba2016,pytorch}. The LSTM receives the sequence of coarse-grained convolutional outputs and models the evolution of latent dynamics. We use a 3-layer LSTM with hidden dimension equal to the number of convolutional outputs (i.e., $N_\ell=192$) to avoid a bottleneck. Moreover, the controller signal depends not just on current inputs but on accumulated energy and system state, i.e., it is non-Markovian. We include two residual skip connections \cite{He2016} over latent features:
\begin{align}\label{eq:skip}
\text{LSTM skip:} \quad & \mathbf h_k^{\text{lstm}} := \text{LSTM}(\mathbf z_k) + \alpha_1 \cdot \text{Linear}(\mathbf z_k), \notag\\
\text{MLP skip:} \quad & \hat{\mathbf p}^\text{LF} := \text{MLP}(\mathbf h_k^{\text{lstm}}) + \alpha_2 \cdot \text{Linear}(\mathbf h_k^{\text{lstm}}),
\end{align}
where $\mathbf z_k\in\mathbb{R}^{N_\ell}$ is the normalized convolutional layer output at each knot interval $k\in\{1,\dots N_k-1\}$, and the scaling parameters \( \alpha_1, \alpha_2 \in [0,2] \) are learned via:
\[
\alpha_i := 2 \cdot \texttt{sigmoid}(\alpha_i^{\text{raw}}).
\]
This allows the network to tune the contribution of early-stage features or shallower representations. Notably, we initialize \( \alpha_1 = 1.0 \) to allow early guidance from the convolutional layer and \( \alpha_2 \approx 0.0 \) to encourage the decoder MLP to learn independently unless the residual improves fit. The final decoder is an MLP \((N_\ell \to N_\ell \to N_\ell \to N_\ell/2 \to 1)\), where each of the three hidden layers are followed by a \texttt{ReLU} activation \cite{pytorch}. Its role is to translate the LSTM’s temporally encoded features into the final output $\hat{\mathbf p}^\text{LF}\in\mathbb{R}^{N_k-1}$. Including multiple fully connected layers, i.e., deepening the decoder, improves the model’s ability to resolve high-curvature regions near transitions (e.g., end-of-dynamics cutoffs) and refine per-time-step predictions without flattening structure \cite[Ch.~6]{goodfellow2016deep}. A summary of the full architecture for the LF surrogate $\F_\text{LF}$ is given in \Cref{fig:LFsurrogate}.

 We train via \texttt{PyTorch} \cite{pytorch} (as are all subsequent NNs) over mini-batches of size $N_B=32$ using the \texttt{Adam} optimizer [ibid] with default parameters \revone{(i.e., \texttt{betas} = (0.9, 0.999))} and an initial learning rate of $5\times10^{-4}$. To prevent instability, we implement gradient clipping with the maximum gradient norm set to 5.0. To improve convergence and prevent overfitting, we employ a \texttt{ReduceLROnPlateau} learning rate scheduler [ibid] based on validation loss with a factor of 0.5 and patience of 50 epochs. Additionally, we apply early stopping with a patience of 300 epochs, which allowed training to converge in approximately $1.1\times10^3$ epochs. 

\section{Dense-time network details}\label{app:inv1}

\begin{figure}[h!]
	\centering  
    \begin{tikzpicture}[
        node distance=1cm and 1cm,
        module/.style={draw, rounded corners, minimum width=2cm, minimum height=.75cm, thick, fill=#1!20},
        arrow/.style={-{Latex[length=3mm]}, thick},
        skip/.style={draw=orange, thick, dashed},
        font=\small
    ]

    \node[draw,rectangle, fill=white!80!black, inner sep=5pt, minimum width=10pt, minimum height=80pt]
    (input) at (0,0) {$\mathbf{x}$};
    \node at ($(input.south)+(0,-0.3)$) {\scriptsize $(N_B, N_t, 2)$};
    
    \node[module=blue] (lstm) [right=of input] {LSTM};
    \node at ($(lstm.south)+(0,-0.3)$) {\scriptsize $(N_B, N_t, N_\ell)$};
    
    \node[module=purple] (attn) [right=of lstm] {Norm \& Pool};
    \node at ($(attn.south)+(0,-0.3)$) {\scriptsize $(N_B, N_\ell)$};
    
    \node[module=green!50!black] (neck) [right=of attn] {Compress};
    \node at ($(neck.north)+(-0.25,+0.3)$) {\scriptsize $(N_B, N_\ell/2)$};

    \node[draw,circle,fill=white,inner sep=1pt] (sum) [right=1.2cm of neck] {$+$};
    
    \node[module=yellow!70!black] (mlp) [above=of sum] {MLP};
    \node at ($(mlp.north)+(0.0,+0.3)$) {\scriptsize $(N_B, N_d)$};

    \node[module=orange!70!black] (skip) [below=of sum] {Skip$_\text{MLP}$};
    \node at ($(skip.north)+(-1.5,+0.4)$) {\scriptsize $\boldsymbol\alpha$};
    \node at ($(skip.south)+(0,-0.3)$) {\scriptsize $(N_B, N_d)$};
    
    \node[draw,rectangle, fill=white!80!black, inner sep=2pt, minimum width=5pt, minimum height=40pt] (output) [right=of sum] {$\hat{\mathbf T}_r^\text{PCA}$};
    \node at ($(output.south)+(0,-0.3)$) {\scriptsize $(N_B, N_d)$};

    \draw[arrow] (input) -- (lstm);
    \draw[arrow] (lstm) -- (attn);
    \draw[arrow] (attn) -- (neck);
    \draw[arrow] (neck) -- (mlp);
    \draw[arrow] (mlp) -- (sum);
    \draw[arrow] (sum) -- (output);

    \draw[skip] (neck) -- (skip);
    \draw[skip] (skip.north) -| (sum);

    \draw[thick] ($(input)+(-1.0,3.1)$) rectangle ($(output)+(1.0,-3.1)$);
    
    \end{tikzpicture}
\caption{Architectural diagram for the inverse model $\mathcal I_\text{D}$, which maps a HF radius and velocity at all $N_t=481$ discrete times to its corresponding static $N_d=4$-dimensional PCA drive representation. Output dimensions for each module are provided, where $N_B$ is the batch size and $N_\ell=32$ is the LSTM hidden dimension.}
\label{fig:InvProb1}
\end{figure}
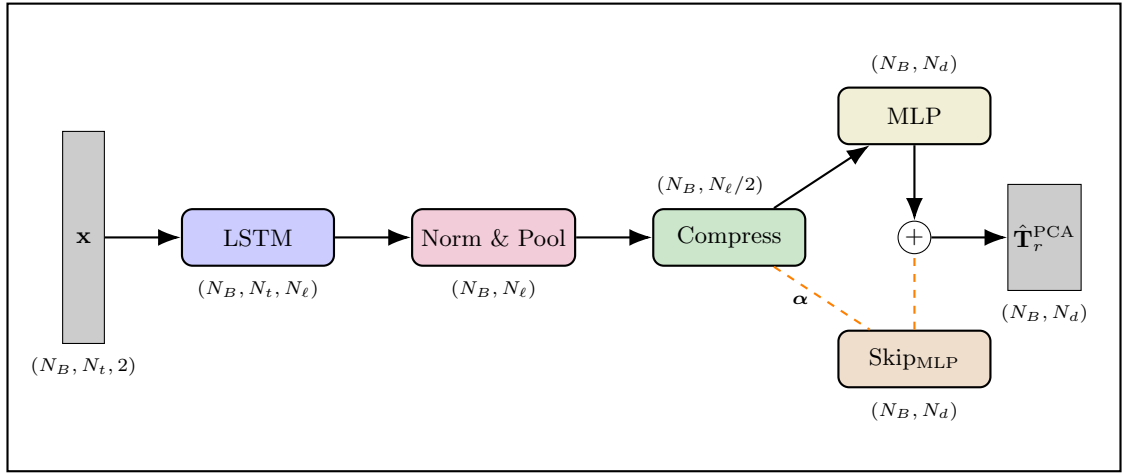

Similar to the motivation behind the HF module of the MF surrogate, the LSTM works well to encode non-local sequential dependencies found in trajectories. Given that the inverse model $\mathcal{I}_\text{D}$ is trained using HF predictions from $\F_\text{MF}$, there is inherent (albeit small) noise in the input training data. Hence, to avoid overfitting, we use a narrower and slightly deeper LSTM encoder; namely, three layers with hidden dimension $N_\ell=32$. The LSTM evolves latent-space features in time via its hidden state $\mathbf{h}_t\in\mathbb{R}^{N_\ell}$ for $t\in\{1,\dots,N_t\}$. To focus on the most informative time steps in the latent space, we learn a scalar score for the normalized hidden state $\mathbf{z}_t\in\mathbb{R}^{N_\ell}$ at each $t\in\{1,\dots,N_t\}$. Specifically, the LSTM output is first passed through a \texttt{LayerNorm} to stabilize gradients, followed by a linear layer of attention queries $\mathbf{w}_q\in\mathbb{R}^{N_\ell}$, which are scored (over time) via the \texttt{SoftMax} operator \cite{pytorch}:
\[
  e_t := \mathbf{w}_q^\top \mathbf{z}_t \in \mathbb{R}
  \quad\Longrightarrow\quad
  \sigma_t := \frac{\exp(e_t)}{\sum_{s=1}^{N_t} \exp(e_s)}\in\mathbb{R},\quad t\in\{1,\dots,N_t\}.
\]
Taking the dot product over time with $\mathbf z_t$ forms a (static) weighted context vector
\[
  \mathbf{c}
  := \sum_{t=1}^{N_t} \sigma_t\,\mathbf{z}_t \in \mathbb{R}^{N_\ell}.
\]
This attention-pooling mechanism enables the model to dynamically select
which temporal features are most relevant for the downstream static regression task \cite{bahdanau2015neural, yang2016hierarchical}. During validation, we found, via PCA, that the context vector's dimensionality could be  reduced from $N_\ell=32$ to $N_\ell/2=16$. Therefore, after passing the context $\mathbf c$ through a \texttt{ReLU} activation, we apply a fully connected feedforward layer (i.e., a $N_\ell\to N_\ell/2$ linear layer and \texttt{ReLU} activation), which serves as a nonlinear bottleneck to improve statistical efficiency and remove redundant dimensions in the pooled latent representation, i.e., to match the LSTM manifold's intrinsic dimensionality. Lastly, the encoded representation is passed through an MLP decoder/prediction head containing three hidden layers ($N_\ell/2\to64\to64\to32\to N_d$) and a gated residual skip connection $(N_\ell/2\to N_d)$, where the latter takes the same form as the skip used in the HF model, i.e., $\alpha := 2 \cdot \texttt{sigmoid}(\alpha^{\text{raw}})$ with $\alpha\approx0$ initialization. This provides high expressivity to map compressed features to PCA outputs, preserving a simple identity path and easing optimization only when necessary to prevent degradation of training performance. An architectural diagram for the network $\mathcal{I}_\text{D}$ is provided in \Cref{fig:InvProb1}. 

\section{Sparse-time network details}\label{app:inv2}

We begin with the model $\mathcal{I}_\text{S}^R$ that uses radius alone. Given an input sample \(\mathbf{R}_{i,n}\in\mathbb{R}^{N_t}\),
we first compute per‐head scores:
\begin{align}\label{eq:score}
  \mathbf{s}^{(n)}
  := W_{\mathrm{score}}\;\mathbf{R}_{i,n}
  \;\in\;\mathbb{R}^{N_s\cdot N_t},
\end{align}
where \(W_{\mathrm{score}}\in\mathbb{R}^{(N_s\cdot N_t)\times N_t}\) is a linear mapping
that outputs \(N_s\) independent score sequences, i.e., $\mathbf{s}^{(n)}$ is reshaped into a matrix in the forward pass.  A scalar parameter \(\gamma > 0\),
learned via
\(\gamma := 2\;\texttt{sigmoid}(\gamma^\text{raw})\in\mathbb{R}\), controls the sharpness of the \texttt{SoftMax}.  We apply \texttt{SoftMax} across time for each head:
\[
  \sigma_{j,t}^{(n)}
  := \frac{\exp\bigl(s_{j,t}^{(n)}/\gamma\bigr)}
         {\sum_{m=1}^{N_t} \exp\bigl(s_{j,m}^{(n)}/\gamma\bigr)}\in\mathbb{R},
  \quad j\in\{1,\dots,N_s\},\; t\in\{1,\dots,N_t\},
\]
yielding a differentiable probability distribution over time steps.  We then form
\(N_s\) soft‐selected features by the weighted summation
\begin{align}\label{eq:soft_times}
  \varphi_{j}^{(n)}
  := \sum_{t=1}^{N_t} \sigma^{(n)}_{j,t}\,R_{i,n}(t)\in\mathbb{R},
  \quad j\in\{1,\dots,N_s\},
\end{align}
producing \(\boldsymbol{\varphi}^{(n)}\in\mathbb{R}^{N_s}\).  The indices
\(\mathbf{I}^{(n)}:=\left\{\,\arg\max_{t}\,\sigma^{(n)}_{j,t}\right\}_{j=1}^{N_s}\in\mathbb{R}^{N_s}\) serve as an interpretable,
discrete approximation of the most important time‐steps for each radius sample $\mathbf{R}_{i,n}$. Lastly, we map \(\boldsymbol{\varphi}^{(n)}\) through an MLP with three hidden layers $(N_s \to 128 \to 128 \to 128 \to N_d)$. 

The inverse model $\mathcal{I}_\text{S}^{R,V}$ that takes in both radius and velocity as input is nearly identical to the model $\mathcal{I}_\text{S}^{R}$ just described. The primary difference is that the input radius $\mathbf{R}_{i,n}$ in \cref{eq:score} now becomes the stacked radius and velocity $\mathbf{x}_n:=(\mathbf{R}_{i,n}^\top,\mathbf{V}_{i,n}^\top)^\top\in\mathbb{R}^{2N_t}$ and the linear score mapping is now $W_\mathrm{score}\in\mathbb{R}^{(N_s\cdot N_t)\times (2N_t)}$, allowing each head to weight both kinematic features dynamically over time. Additionally, each of the $N_s$ soft-selected features in \cref{eq:soft_times} are now 2D vectors $\boldsymbol{\varphi}_j^{(n)}\in\mathbb{R}^2$ due to the radius $R_{i,n}(t)\in\mathbb{R}$ being replaced by $\x_n(t)\in\mathbb{R}^2$, which results in the feature vector becoming a feature matrix \(\phi^{(n)}\in\mathbb{R}^{N_s\times 2}\) corresponding to radius and velocity. Lastly, we add an additional hidden layer of equal width to the MLP (i.e., resulting in four hidden layers each with 128 neurons) with a single dropout layer of $5\%$ placed directly in the middle of the MLP. This additional layer helps to compensate for the slightly more complex nonlinear map resulting from doubling the MLP input dimension to account for velocity (i.e., the feature matrix \(\phi^{(n)}\) is flattened before being passed to the MLP), and the dropout layer helps prevent overfitting of edge cases \cite{Srivastava2014}, reducing the tail of the error distributions.

\end{document}